\documentclass[aps,prd,twocolumn,preprintnumbers,superscriptaddress,amssymb,eqsecnum,showpacs,showkeyes,secnumarabic,graphics,floatfix,nofootinbib,tightenlines,longbibliography]{revtex4-1}

\usepackage{natbib}

\usepackage{graphicx}
\usepackage{bm}
\usepackage{cancel}
\usepackage{epsfig}
\usepackage{dcolumn}
\usepackage{amsmath}
\usepackage{hhline}
\usepackage{enumerate}
\usepackage[utf8]{inputenc}
\usepackage{ulem}
\usepackage[dvipsnames]{xcolor}
\usepackage[breaklinks=true,colorlinks=true,
linkcolor=blue,urlcolor=Blue,citecolor=MidnightBlue,
bookmarks=true,bookmarksopenlevel=2]{hyperref}
\usepackage{bm}
\usepackage{amsmath}
\usepackage{amssymb}
\usepackage{longtable}
\usepackage{breakcites}
\usepackage{color, colortbl}
\usepackage{appendix}
\definecolor{LightGreen}{rgb}{0.88,1,0.88}
\definecolor{LightCyan}{rgb}{0.88,1,1}
\definecolor{LightRed}{rgb}{1,0.85,0.85}

\def\GA#1{\textcolor{Red}{#1}}

\newcommand{\ie}{{\it i.e.} }

\begin{document}
\makeatother

\preprint{IFT-UAM/CSIC-24-2}

\title{Applying the Viterbi Algorithm to Planetary-Mass Black Hole Searches}

\author{George Alestas}\email{g.alestas@csic.es}
\affiliation{Instituto de F\'isica Te\'orica UAM-CSIC, Universidad Auton\'oma de Madrid,
Cantoblanco, 28049 Madrid, Spain}
\author{Gonzalo Morr\'as}\email{gonzalo.morras@uam.es}
\affiliation{Instituto de F\'isica Te\'orica UAM-CSIC, Universidad Auton\'oma de Madrid,
Cantoblanco, 28049 Madrid, Spain}
\author{Takahiro S. Yamamoto}\email{yamamoto.takahiro.u6@f.mail.nagoya-u.ac.jp}
\affiliation{Department of Physics, Nagoya University, Nagoya, 464-8602, Japan}
\author{Juan Garc\'ia-Bellido}\email{juan.garciabellido@csic.es}
\affiliation{Instituto de F\'isica Te\'orica UAM-CSIC, Universidad Auton\'oma de Madrid,
Cantoblanco, 28049 Madrid, Spain}
\author{Sachiko Kuroyanagi}\email{sachiko.kuroyanagi@csic.es}
\affiliation{Instituto de F\'isica Te\'orica UAM-CSIC, Universidad Auton\'oma de Madrid,
Cantoblanco, 28049 Madrid, Spain}
\affiliation{Department of Physics, Nagoya University, Nagoya, 464-8602, Japan}
\author{Savvas Nesseris}\email{savvas.nesseris@csic.es}
\affiliation{Instituto de F\'isica Te\'orica UAM-CSIC, Universidad Auton\'oma de Madrid,
Cantoblanco, 28049 Madrid, Spain}

\date{\today}

\begin{abstract}
\noindent 
The search for subsolar mass primordial black holes (PBHs) poses a challenging problem due to the low signal-to-noise ratio, extended signal duration, and computational cost demands, compared to solar mass binary black hole events. In this paper, we explore the possibility of investigating the mass range between subsolar and planetary masses, which is not accessible using standard matched filtering and continuous wave searches. We propose a systematic approach employing the Viterbi algorithm, a dynamic programming algorithm that identifies the most likely sequence of hidden Markov states given a sequence of observations, to detect signals from small mass PBH binaries. We formulate the methodology, provide the optimal length for short-time Fourier transforms, and estimate sensitivity. Subsequently, we demonstrate the effectiveness of the Viterbi algorithm in identifying signals within mock data containing Gaussian noise. Our approach offers the primary advantage of being agnostic and computationally efficient.

\end{abstract}
\maketitle

\section{Introduction}
\label{sec:intro}

The study of black holes has become a central focus, driven by the detection of gravitational waves (GWs) resulting from binary black hole mergers~\cite{LIGOScientific:2016aoc, LIGOScientific:2018mvr, LIGOScientific:2020ibl, LIGOScientific:2021usb, LIGOScientific:2021djp}. Simultaneously, there is a growing interest in primordial black holes (PBHs) — black holes formed in the early universe — as a potential explanation for the origin of observed black holes~\cite{Bird:2016dcv, Clesse:2016vqa, Sasaki:2016jop, Clesse:2017bsw, Clesse:2020ghq, Carr:2023tpt}, some of which pose challenges for explanations based on astrophysical origins. The concept of PBH was initially speculated by Zeldovich and Novikov~\cite{1967SvA....10..602Z} and later formally proposed by Hawking and Carr~\cite{Hawking:1971ei,Carr:1974nx, Carr:1975qj} (see also~\cite{Garcia-Bellido:1996mdl,  Carr:2016drx,Carr:2020xqk,Escriva:2022duf,LISACosmologyWorkingGroup:2023njw}). PBHs are also fascinating as a strong candidate for providing a possible explanation for dark matter. Depending on the time of their formation, their masses can vary from the Planck mass to the mass of a supermassive black hole~\cite{Carr:2009jm, Carr:2020gox}.

A detection of a subsolar mass black hole by the LIGO-Virgo-KAGRA (LVK) collaboration \cite{LIGOScientific:2014pky, VIRGO:2014yos, PhysRevD.88.043007, abbott2020prospects} would provide smoking gun evidence for PBH, as astrophysical black holes are expected to have masses larger than $\sim 1 M_\odot$. Searches for subsolar mass binary black hole events have been performed with the first (O1), second (O2), and third (O3) observing run data~\cite{LIGOScientific:2018glc,LIGOScientific:2019kan,Nitz:2021mzz, Nitz:2022ltl,LIGOScientific:2021job,LIGOScientific:2022hai,Phukon:2021cus,Morras:2023jvb} using the matched filtering method for compact binary coalescence (CBC), and no significant event has been identified so far. The primary challenge in subsolar mass searches is computational time, as the number of templates and their duration increases dramatically when the minimum mass included in the search is small~\cite{Magee:2018opb}, limiting the current search range of the smallest black hole mass to be $0.1 M_\odot$. For smaller masses, the signal becomes more similar to continuous waves (CWs), and constraints on planetary-mass and asteroid-mass have been reported using the upper limits obtained by CW searches with LVK O3 data~\cite{Miller:2021knj, KAGRA:2022dwb}.  In this case, the frequency evolution of the inspiral signal must be smaller than the maximum spin-up allowed in the CW search (typically $\dot{f} < 10^{-9}$~Hz/s). This corresponds to an upper mass limit of $< 3 \times 10^{-5}M_\odot$ for equal mass binaries.

In this paper, we aim to investigate the possibility of exploring the intermediate mass range that is not covered by the subsolar CBC and CW searches, specifically $3 \times 10^{-5}M_\odot < M < 0.1 M_\odot$. We still employ a CW search algorithm, but the segment size of the short-time Fourier transform (SFT) must be optimally adjusted depending on the frequency evolution of the signal, which is determined by the chirp mass. In this case, the SFT length is much shorter than those used in typical
CW searches. It is worth noting that the investigation into this mass range complements microlensing observations and helps to explore the mass range indicated by ultrashort-timescale OGLE events ($10^{-6} - 10^{-4} M_\odot$)~\cite{Niikura:2019kqi}.

The pioneering work in exploring this mass range with GW data~\cite{Miller:2020kmv} is developed assuming the Frequency-Hough and Generalized Frequency-Hough algorithms, primarily designed for CW searches. In our study, we employ the Viterbi dynamic programming algorithm~\cite{Viterbi}, which identifies the most likely sequence of hidden Markov states in a model-agnostic way. The application of the Viterbi statistic to CW searches has been extensively studied~\cite{Suvorova:2016rdc,Suvorova:2017dpm,Melatos:2021mmz,Bayley:2019bcb, Bayley:2020zfa,Bayley:2022hkz,KAGRA:2022dwb} (see also~\cite{Sun2018,Sun:2018owi,Melatos:2020rlu} for other applications), and the full search pipeline is available as the SOAP package~\cite{soapcw}. As observed in the CW searches~\cite{KAGRA:2022dwb}, we expect the Viterbi algorithm to be less sensitive compared to the Frequency-Hough algorithm, but it offers significant advantages in terms of computation time and an agnostic approach. The Frequency-Hough method assumes a power-law spectrum, while Viterbi, in principle, can detect arbitrary curves in the time-frequency domain. This flexibility is a major advantage in the portion of the parameter space where higher-order post-Newtonians (PNs) start to become non-negligible.

The structure of the paper is the following: In Sec.~\ref{sec:meth}, we first present a summary of the Viterbi algorithm and discuss its theoretical basis. Subsequently, we show that the length of the SFT needs to be optimized based on the chirp mass of the binary system. Then, we provide a detailed description of how to compute the optimum length.
We also provide an estimation of the sensitivity in terms of distance and its implications for the PBH abundance that could be obtained through this search. Additionally, we discuss the detection statistics with a detailed investigation into their distribution. 
Following that, in Sec.~\ref{sec:results}, we demonstrate the application of SOAP package to the search of planetary mass binary black holes by preparing mock
data containing Gaussian noise and injections of signals, with varying luminosity distances and chirp masses.
Finally, in Sec.~\ref{sec:Conclusions}, we present the conclusions of our study and the outlook for future investigations. Appendix~\ref{sec:LOPN_val} demonstrates that higher-order PNs are indeed non-negligible for most of the parameter space, supporting the advantage of the model-agnostic nature of our method.

\section{Methodology}
\label{sec:meth}
\subsection{Viterbi Algorithm Framework}
\label{subsec:theory}

In this work, we make use of the Viterbi algorithm, via the SOAP package~\cite{Bayley:2019bcb, Bayley:2020zfa, Bayley:2022hkz, KAGRA:2022dwb}, in order to search for hidden GW signals belonging to planetary mass black hole mergers. The Viterbi algorithm is a dynamic programming algorithm that is used to find the most probable sequence of hidden states in a Markov model that is based on data. It focuses on maximizing the probability of a signal existing within the data in every possible direction, after each discrete step in the detection process. A significant merit of this process is the fact that less computational cost is needed since there is a significant reduction in the number of possible tracks that need to be calculated beforehand.

In the SOAP algorithm, the time series data is divided into $N$ segments of equal length, forming the dataset of time series segments $\bm{x}_i$, denoted as $D\equiv \{ \bm{x}_i \}$, where $i$ labels the time sequence of segments. The track we are looking for is a list of frequencies $\bm{\nu} \equiv \{ \nu_i \}$, where $\nu_i$ is the frequency of the GW signal in the segment $\bm{x}_i$.
The main goal of the algorithm is to maximize the posterior probability of each possible signal track in order to identify the most probable one. Following the Bayes theorem, the posterior probability takes the form, 
\begin{equation}
    p({\bm \nu} \mid D) = \frac{p({\bm \nu})p(D \mid {\bm \nu})}{p(D)}, \label{eq:bayes_post_prob}
\end{equation}
where $p(D \mid {\bm \nu})$ is the likelihood, or probability of a signal existing within the track, $p({\bm \nu})$ is the prior probability of the track and $p(D)$ is the marginalized likelihood of the model. We can split the prior probability of the track in a set of transition probabilities, 
\begin{align}
p({\bm \nu}) &= p(\nu_{N - 1}, \ldots, \nu_1, \nu_0)\nonumber \\
&= p(\nu_0)\prod_{i=1}^{N-1}p(\nu_i \mid \nu_{i-1}),
\label{eq:pnu}
\end{align}
where $\nu_0$ is the frequency of the first time-step and $p(\nu_i \mid \nu_{i-1})$ is the transition probability for $\nu_i$ given the frequency at the previous time-step was $\nu_{i-1}$.
We use the same transition probability as in~\cite{Bayley:2019bcb}.

In this light, the posterior Bayesian probability characterized by Eq. \eqref{eq:bayes_post_prob} takes the form, 
\begin{equation}
\label{eq:post_prob}
    p({\bm \nu} | D) =
    \frac{p(\nu_0)p({\bm x_0} | \nu_0) \displaystyle\prod_{i=1}^{N-1}p(\nu_i
| \nu_{i-1})p({\bm x_i} | \nu_i)}{\displaystyle\sum_{S}
\left\{p(\nu_0)p({\bm x_0} | \nu_0)\displaystyle\prod_{i=1}^{N-1}p(\nu_i |
\nu_{i-1})p({\bm x_i} | \nu_i)\right\}} \,,
\end{equation}
where we have used the fact
that we can factorize the likelihood $p(D \mid {\bm \nu})$ as 
\begin{equation}
p(D \mid {\bm \nu}) = \prod_{i=0}^{N-1}p({\bm x_i} \mid \nu_i).
\end{equation}
The most probable signal track is then found by maximizing the logarithm of the numerator in Eq. \eqref{eq:post_prob}
\begin{equation}
    \begin{split}
        \log p(\hat{\bm \nu} | D)  = \max_{{\bm \nu}}{\biggl\{ \log p(\nu_0) + \log p({\bm x_0} | \nu_0)  } \\ 
        \left. + \sum_{i=1}^{N-1} \biggl[ \log p(\nu_i | \nu_{i-1}) + \log p({\bm x_i} | \nu_i) \biggr] \right\} + \text{const.} \,,\label{eq:maxtracklog}
    \end{split}
\end{equation}
for each frequency at each time step.

\subsection{Short-Time Fourier Transforms}
\label{sec:SFT_theory}

The main input of the Viterbi algorithm
is the Short-Time Fourier Transforms (SFTs) of the available data frames. More specifically, the SFT computation process works by dividing the time function $s[n]$ of the strain data into smaller chunks, typically 
using a rectangular window function $w[n]$, where $n$ is the label for a time series of discrete data points. With a finite-duration window from $0$ to $N_w -1$, the SFT is given by
\begin{equation}
    \tilde{s}[n, k] = \sum^{n}_{m=n-(N_w-1)} w[m-n]s[m]e^{-i \frac{2 \pi k}{N_f} m }\,,
\end{equation}
where $N_f = T_\mathrm{SFT} f_\mathrm{s}/2$, with $f_\mathrm{s}$ being the sampling frequency, is the number of discrete frequency channels and 
$k$ is the label for GW frequency bins. Here, the variable $m$ is a dummy time index, and $n$ pinpoints the location of the window along the time axis.

In order to apply continuous wave search techniques, our main assumption is that, in each of the SFTs, the signal has to be contained in a single frequency bin. Therefore, in $i$-th SFTs, the waveform can be approximated as
\begin{equation}
    h(t) = h_{i} \cos({2 \pi f_i t + \phi_i}) \rightarrow \tilde{h}(f) = \frac{1}{2} h_i e^{i \phi_i} \delta(f - f_i) \,.
    \label{eq:h_i}
\end{equation}
In the case of a compact binary coalescence, the time at which the signal has a frequency $f$ is given by~\cite{Maggiore_Vol1},
\begin{equation}
    t_\mathrm{GW}(f) = t_c - \frac{5}{256} \left(\frac{G \mathcal{M}_c}{c^3} \right)^{-5/3} (\pi f)^{-8/3} \,,
    \label{eq:t_f_CBC}
\end{equation}
which is a monotonically increasing function. 
Here, $t_c$ is the time of coalescence and $\mathcal{M}_c$ is the chirp mass.
The frequency bin size is determined by $\Delta f = 1/T_{\rm SFT}$, where $T_{\rm SFT}$ is the SFT length. Thus, the condition that the signal stays in the same frequency bin (between $f-\Delta f/2$ and $f+\Delta f/2$) throughout the SFT time is given by
\begin{equation}
    T_\mathrm{SFT} \leq t_\mathrm{GW}(f + \frac{1}{2T_\mathrm{SFT}}) - t_\mathrm{GW}(f- \frac{1}{2T_\mathrm{SFT}})\,,
    \label{eq:MonochromaticCondition}
\end{equation}
while the maximum frequency $f_*$ for which the inequality is satisfied is given by
\begin{align}
    T_\mathrm{SFT} & = t_\mathrm{GW}(f_* + \frac{1}{2T_\mathrm{SFT}}) - t_\mathrm{GW}(f_* - \frac{1}{2T_\mathrm{SFT}})\nonumber\\
    & \approx \frac{1}{T_\mathrm{SFT}} \frac{d t_\mathrm{GW}}{d f} = \frac{1}{T_\mathrm{SFT}} \frac{5 \pi}{96} \left(\frac{G \mathcal{M}_c}{c^3} \right)^{-5/3} (\pi f_*)^{-11/3}.
    \label{eq:MonochromaticCondition_maxf}    
\end{align}
Finally, solving for $f_*$, we obtain
\begin{align}
    f_* &= \frac{1}{\pi} \left(\frac{5 \pi}{96}\right)^{3/11} \left(\frac{G \mathcal{M}_c}{c^3} \right)^{-5/11} T_\mathrm{SFT}^{-6/11} .
    \label{eq:f_star}
\end{align}

\subsection{Optimum SFT length}
\label{sec:opt_TSFT_constw}

In order to find the optimum SFT length, let us define the total signal-to-noise ratio (SNR) given by adding the SNR of each of these segments
\begin{equation}
    \rho^2_\mathrm{tot} = \sum_i |\rho^\mathrm{mf}_i|^2 \,.
\label{eq:tot_SNR}
\end{equation}
Here, $\rho^\mathrm{mf}_i$ is the matched filter SNR in the $i$-th SFT,
\begin{equation}
    \rho^\mathrm{mf}_i = \frac{\langle h , s_i \rangle}{\rho^\mathrm{opt}_i} \,,
    \label{eq:rho_i}
\end{equation}
where $h$ and $s_i$ represent the GW template and the strain data of i-th segment, respectively,
and $\langle \cdot , \cdot \rangle$ represents the noise-weighted inner product, defined as
\begin{equation}
    \langle a, b \rangle = 4 \int_{f_{\mathrm{min}}}^{f_{\mathrm{max}}} \frac{\tilde{a}^{*}(f)\tilde{b}(f)}{S_{n}(f)} df \,,
    \label{eq:inner_product}
\end{equation}
where the tilde denotes the Fourier transform. The optimum SNR $\rho^\mathrm{opt}_i$ is given by
\begin{equation}
    \rho^\mathrm{opt}_i = \sqrt{\langle h , h \rangle},
    \label{eq:opt_snr}
\end{equation}
where $S_{n}(f)$ is the noise power spectral density (PSD) of the detector. When the signal is given by 
the monochromatic form as in Eq.~\eqref{eq:h_i} in each SFT,
the optimum SNR, defined in Eq.~\eqref{eq:opt_snr}, is approximately equal to
\begin{equation}
    \rho^\mathrm{opt}_i \simeq \sqrt{\frac{T_\mathrm{SFT}}{S_{n}(f_i)}} h_i,
    \label{eq:opt_snr_cw}
\end{equation}
and the matched filter SNR, defined in Eq.~\eqref{eq:rho_i}, is given by
\begin{equation}
    \rho^\mathrm{mf}_i \simeq \frac{2 \tilde{s}(f_i) e^{i \phi_i}}{\sqrt{S_{n}(f_i) T_\mathrm{SFT}}}.
    \label{eq:mf_snr_cw}
\end{equation}

Assuming that the detector strain $\tilde{s}(f_i)$ has a signal part $h_i$ and a Gaussian noise part $n_i$, the matched filter SNR in each SFT of Eq.~\eqref{eq:mf_snr_cw} is a complex normal variable with a mean given by the optimum SNR of Eq.~\eqref{eq:opt_snr_cw}. Therefore, the square of the total incoherent SNR of Eq.~\eqref{eq:tot_SNR} is distributed like a non-central $\chi^2$ distribution with $2 N_\mathrm{SFT}$ degrees of freedom and non-centrality parameter $(\rho^\mathrm{opt}_\mathrm{tot})^2$, 
\begin{equation}
    p(x) = \frac{1}{2} e^{-(x + (\rho^\mathrm{opt}_\mathrm{tot})^2)/2} \left( \frac{x}{(\rho^\mathrm{opt}_\mathrm{tot})^2} \right)^{(N_\mathrm{SFT} -1)/2} I_{N_\mathrm{SFT} -1}(\rho^\mathrm{opt}_\mathrm{tot}\sqrt{x}),
    \label{eq:x_dist}
\end{equation}
where $N_\mathrm{SFT} = T/T_\mathrm{SFT}$ is the number of SFTs with $T$ being the total observation time, $I_\nu$ is a modified Bessel function of the first kind, and $\rho^\mathrm{opt}_\mathrm{tot}$ is the total optimal SNR 
obtained by summing over optimal SNRs of those bins in which the signal is monochromatic
\begin{equation}
    \rho^\mathrm{opt}_\mathrm{tot} = \sqrt{4 \int_{f_0}^{f_* (T_\mathrm{SFT})} df \frac{|\tilde{h}(f)|^2}{S_n (f)}}.
    \label{eq:rho_opt_tot}
\end{equation}
The probability distribution of Eq.~\eqref{eq:x_dist} has the following mean $\mu$ and standard deviation $\sigma$~\cite{Abramowitz_and_Stegun}
\begin{subequations}
\label{eq:mu_and_sigma_x_dist}
\begin{align}
    \mu(\rho^\mathrm{opt}_\mathrm{tot}, N_\mathrm{SFT}) & = 2 N_\mathrm{SFT} + (\rho^\mathrm{opt}_\mathrm{tot})^2 \; , \label{eq:mu_and_sigma_x_dist:mu}\\
    \sigma(\rho^\mathrm{opt}_\mathrm{tot}, N_\mathrm{SFT}) & = 2 \sqrt{ N_\mathrm{SFT} + (\rho^\mathrm{opt}_\mathrm{tot})^2} \; . \label{eq:mu_and_sigma_x_dist:sigma}
\end{align}
\end{subequations}

For very large $N_\mathrm{SFT}$, the higher order moments tend to 0, and as expected from the central limit theorem, the distribution of Eq.~\eqref{eq:x_dist} can be approximated by a Gaussian with the mean and standard deviation given in Eq.~\eqref{eq:mu_and_sigma_x_dist}. Assuming that $N_\mathrm{SFT}\gg 1$ and a Gaussian is a good approximation, 
the number of ``sigmas'' $n_\sigma$ at which a signal with optimum SNR $\rho^\mathrm{opt}_\mathrm{tot}$ is expected to be is given by
\begin{equation}
    n_\sigma = \frac{\mu(\rho^\mathrm{opt}_\mathrm{tot}, N_\mathrm{SFT}) - \mu(0, N_\mathrm{SFT})}{\sigma(0, N_\mathrm{SFT})} = \frac{(\rho^\mathrm{opt}_\mathrm{tot})^2}{2 \sqrt{ N_\mathrm{SFT}}} \;.
    \label{eq:n_sigmas}
\end{equation}
So if we want to detect a signal with a significance of $n_\sigma$, it needs to have an optimum SNR larger than
\begin{equation}
    \rho^\mathrm{opt}_\mathrm{tot} = (4 n_\sigma^2 N_\mathrm{SFT})^{1/4}\,.
    \label{eq:rho_n_sigmas}
\end{equation}

To optimize our search, we want to maximize $n_\sigma$. 
For the inspiral part of a CBC signal, using the stationary phase approximation and 
multiply the factor $(2/5)^2$ to average over the polarization, position in the sky, and inclination, the squared amplitude of the frequency domain strain is given by~\cite{Maggiore_Vol1}
\begin{equation}
    |\tilde{h}(f)|^2 = \frac{1}{30 \pi^{4/3}} \left(\frac{c}{d_L}\right)^2 \left(\frac{G \mathcal{M}_c}{c^3}\right)^{5/3} f^{-7/3} \propto f^{-7/3} ,
    \label{eq:CBC_fd_amplitude}
\end{equation}
where $d_L$ is the luminosity distance to the source. Using $N_\mathrm{SFT}=T/T_\mathrm{SFT}$, Eq.~\eqref{eq:n_sigmas} can be rewritten as
\begin{align}
    n_\sigma & = \frac{\left(c/d_L\right)^2 \left(G \mathcal{M}_c/c^3\right)^{5/3}}{15 \pi^{4/3}} \sqrt{\frac{T_\mathrm{SFT}}{T}}\int_{f_0}^{f_*}  \frac{df}{f^{7/3} S_n (f)}.\nonumber \\
    \label{eq:n_sigmas_TSFT}    
\end{align}

\begin{figure}[!t]
\begin{center}
\includegraphics[width=0.5\textwidth]{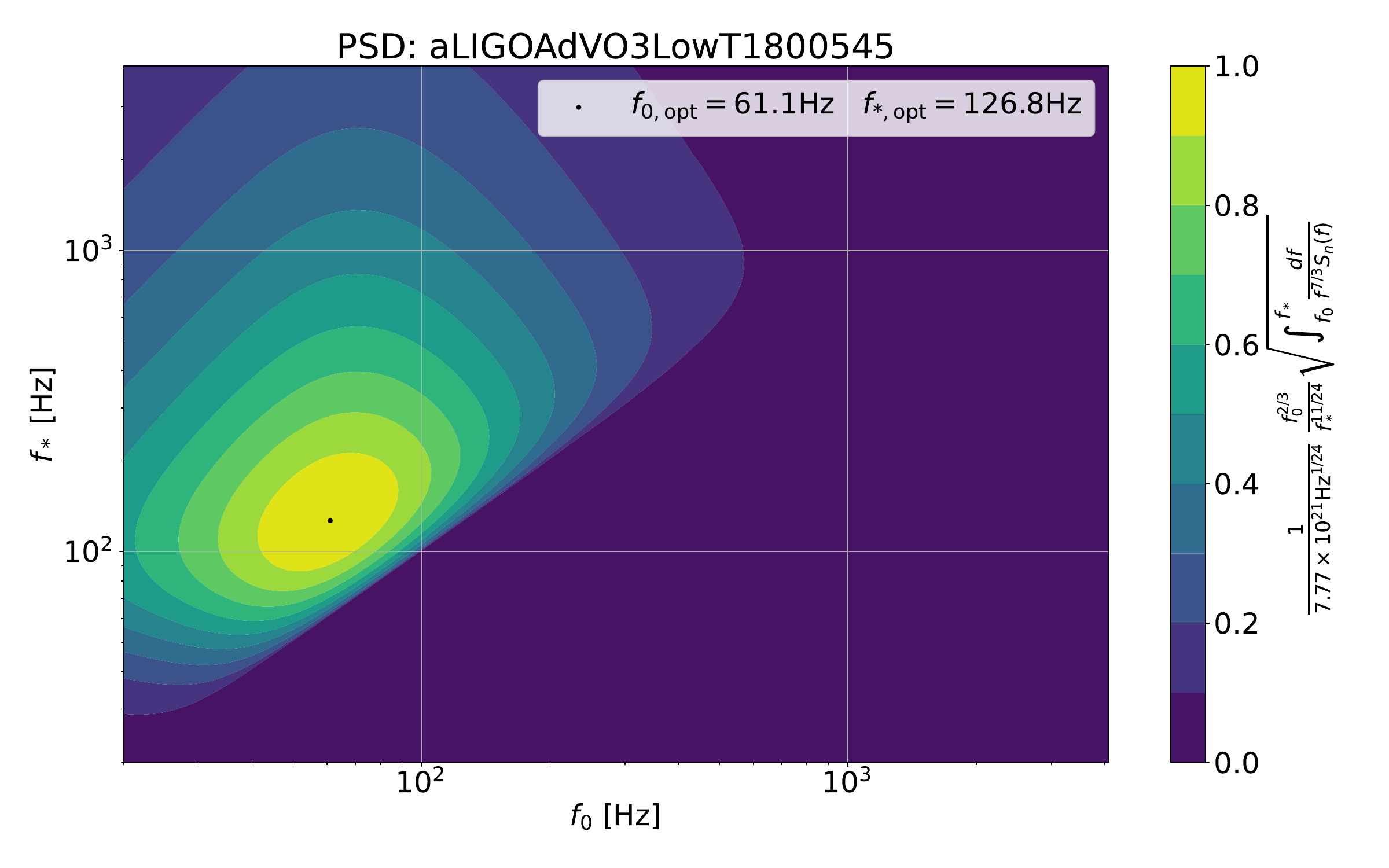}
\end{center} 
\caption{Colormap of $F(f_0,f_*)$ (defined in Eq.~\eqref{eq:F_f0fs}) normalized by its maximum value as a function $f_0$ and $f_*$. For $S_n(f)$ we assume a typical LIGO O3 PSD~\cite{lalsuite}. We also show with a dot the point where the maximum of the function is reached, for $f_{0, \mathrm{opt}} = 61.1$Hz and $f_{*, \mathrm{opt}} = 126.8$Hz with a value of $F(f_{0, \mathrm{opt}},f_{*, \mathrm{opt}}) = 7.77 \times 10^{21}\mathrm{Hz}^{1/24}$}. 
\label{fig:f0fs_colormap}
\end{figure}

\begin{figure}[!t]
\begin{center}
\includegraphics[width=0.5\textwidth]{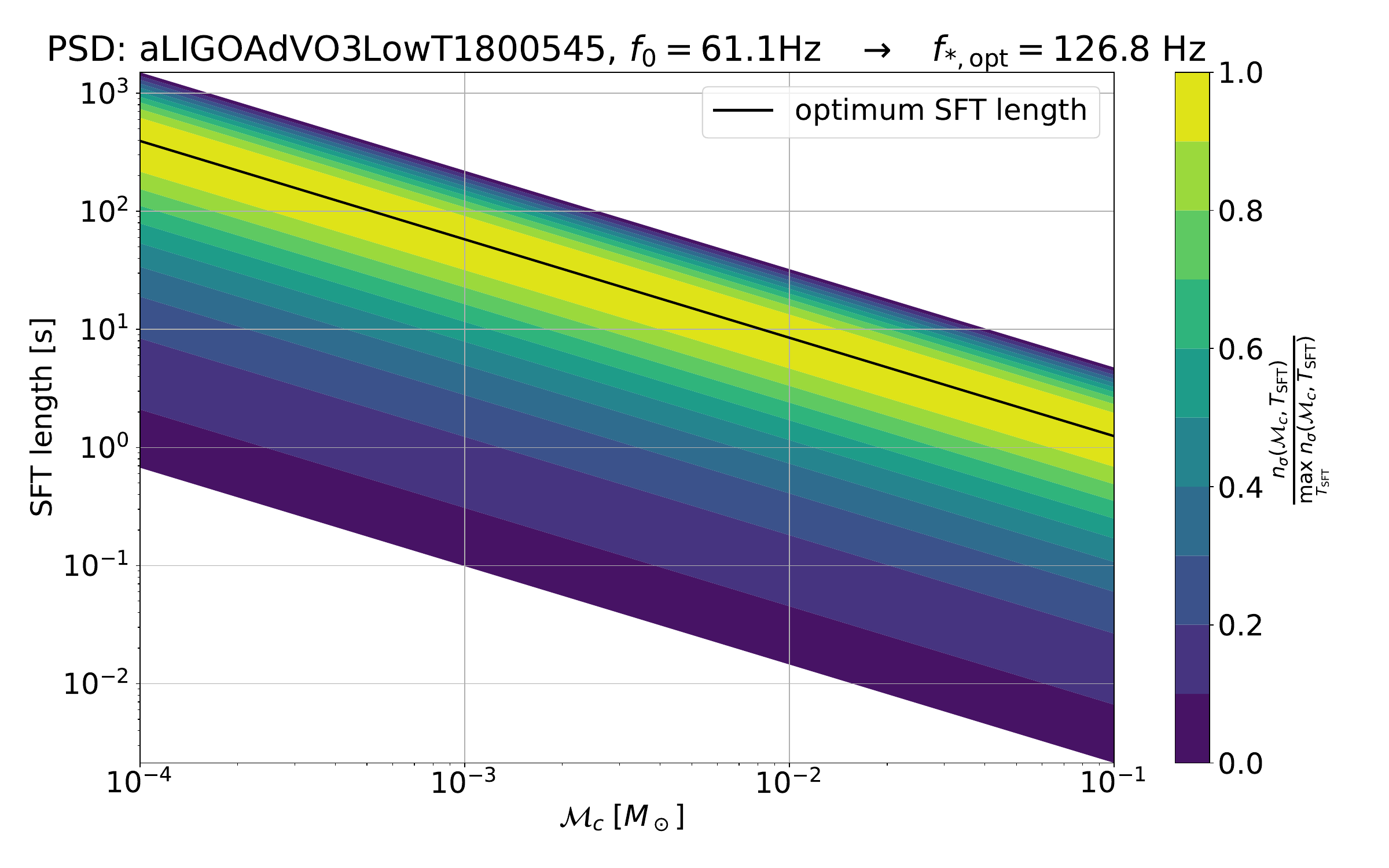}
\caption{\label{plt:SFT_colormap}Colormap of $n_\sigma$ given by Eq.~\eqref{eq:n_sigmas}, normalized to the maximum possible $n_\sigma$ that can be obtained at each chirp mass as a function of the SFT length and the chirp mass. We also show with a solid black line the optimum SFT length as a function of the chirp mass. We assume the same typical LIGO O3 PSD of Fig.~\ref{fig:f0fs_colormap} and use the optimum low-frequency cutoff of $f_0 = 61.1\mathrm{Hz}$.}
\end{center} 
\end{figure}

To assess the potential of our search, it is interesting to compute its range of reach as a function of the chirp mass, the SFT length, and the significance $n_\sigma$. This can be done by solving for $d_L$ in Eq.~\eqref{eq:n_sigmas_TSFT}, 
\begin{equation}
    \frac{d_L}{c} = \sqrt{\frac{ \left(G \mathcal{M}_c/c^3\right)^{5/3}}{15 \pi^{4/3} n_\sigma} \sqrt{\frac{T_\mathrm{SFT}}{T}}\int_{f_0}^{f_* (T_\mathrm{SFT})} \frac{df}{f^{7/3} S_n (f)}}.
    \label{eq:horizon_distance}
\end{equation}
Here, the observation time $T$ that needs to be substituted into this equation is the time when the data and the CBC signal coincide:
\begin{equation}
    T = T(\mathrm{data} \cap \mathrm{CBC \; signal}) \,,
    \label{eq:T_horizon_distance}
\end{equation}
and the integral of Eq.~\eqref{eq:horizon_distance} has to be computed only over the frequencies contained in this time.
The length of the CBC signal between the low-frequency cutoff $f_0$ and the frequency $f_*$, at which it stops being contained in a single SFT, can be obtained from Eq.~\eqref{eq:t_f_CBC},
\begin{align}
    T_\mathrm{CBC} & = \frac{5}{256} \left(\frac{G \mathcal{M}_c}{c^3} \right)^{-5/3}\left[ (\pi f_0)^{-8/3} - (\pi f_*)^{-8/3}\right] \nonumber \\
    & \approx \frac{5}{256} \left(\frac{G \mathcal{M}_c}{c^3} \right)^{-5/3} (\pi f_0)^{-8/3},
    \label{eq:CBC_duration}    
\end{align}
where we have used the approximation $(\pi f_0)^{-8/3} \gg (\pi f_*)^{-8/3}$. For simplicity, we assume that the signal is fully contained in the data and so $T = T_\mathrm{CBC}$. Note that, using a low-frequency cutoff of 60Hz, CBC signals with chirp masses smaller than $1.4 \times 10^{-4} M_\odot$ lasts more than one year in the detector, as we can see from Eq.~\eqref{eq:CBC_duration}.

Substituting $T = T_\mathrm{CBC}$ in Eq.~\eqref{eq:horizon_distance}, we obtain
\begin{equation}
    \frac{d_L}{c} = \left( \frac{256 T_\mathrm{SFT}}{1125 n_\sigma^2}\right)^{1/4} f_0^{2/3} \left(\frac{G \mathcal{M}_c}{c^3}\right)^{5/4} \sqrt{\int_{f_0}^{f_*} \frac{df}{f^{7/3} S_n (f)}},
    \label{eq:horizon_distance_TCBC}
\end{equation}
and using the relation between  $T_\mathrm{SFT}$ and $f_*$ of Eq.~\eqref{eq:f_star}, we 
obtain
\begin{equation}
    \frac{d_L}{c} = \frac{2^{11/8} \left(G \mathcal{M}_c/c^3 \right)^{25/24}}{15^{5/8} \pi^{1/3}} \frac{f_0^{2/3}}{f_*^{11/24}}\sqrt{\frac{1}{n_\sigma}\int_{f_0}^{f_*} \frac{df}{f^{7/3} S_n (f)}}.
    \label{eq:horizon_distance_TCBC_f0_fs}
\end{equation}

Here, $f_0$ and $f_*$ are parameters that can be tuned by changing the low-frequency cutoff and the SFT length. In practice, we want to choose the values of $f_0$ and $f_*$ that maximize the luminosity distances that our searches can reach.
In Eq.~\eqref{eq:horizon_distance_TCBC_f0_fs}, we can separate the part that depends on $f_0$ and $f_*$, and the function that we want to maximize to obtain the optimum $f_0$ and $f_*$ is
\begin{equation}
    F(f_0, f_*) = \frac{f_0^{2/3}}{f_*^{11/24}} \sqrt{\int_{f_0}^{f_*} \frac{df}{f^{7/3} S_n (f)}}.
    \label{eq:F_f0fs}
\end{equation}
Note that this function is solely determined by the shape of the PSD and does not depend on the signal parameters, such as the chirp mass. In Fig.~\ref{fig:f0fs_colormap}, we show the colormap of $F(f_0, f_*)$ computed for a typical LIGO O3 PSD~\cite{abbott2020prospects}, where we observe that there is a well-defined maximum at $f_{0, \mathrm{opt}} = 61.1$Hz and $f_{*, \mathrm{opt}} = 126.8$Hz. To obtain this value of $f_*$, the optimum SFT length depends on the chirp mass as given in Eq.~\eqref{eq:f_star}
\begin{align}
    T_\mathrm{SFT}^\mathrm{opt} & =  \left(\frac{5 \pi}{96}\right)^{1/2} \left(\frac{G \mathcal{M}_c}{c^3} \right)^{-5/6} (\pi f_{*, \mathrm{opt}})^{-11/6} 
    \nonumber \\  & = 8.50 ~ \mathrm{s} \left(\frac{\mathcal{M}_c}{10^{-2} M_\odot}\right)^{-5/6} \left(\frac{f_{*, \mathrm{opt}}}{126.8 ~\mathrm{Hz}}\right)^{-11/6}.
    \label{eq:TSFTopt}    
\end{align}
In Fig.~\ref{plt:SFT_colormap}, we plot the optimum SFT length as a function of chirp mass. We also plot the colormap of $n_\sigma/n_{\sigma,{\rm max}}$, where $n_{\sigma,{\rm max}}$ is the value obtained when we apply the optimal SFT length. This shows how much the significance gets degraded when we apply non-optimum SFT length.

Then, using the maximum of $F(f_0, f_*)$, we can derive the maximum distance at which we can see an event as a function of the chirp mass and $n_\sigma$
\begin{align}
    d_L \! = \! 0.190\,\mathrm{Mpc} \!\!\left(\frac{\mathcal{M}_c}{10^{-2} M_\odot} \right)^{25/24} \! \! \left(\frac{n_\sigma}{10} \right)^{-1/2} \!\! \frac{F(f_0, f_*)}{7.77 \times 10^{21} \mathrm{Hz}^{1/24}}.
    \label{eq:dL_numbers}
\end{align}
Thus, for chirp masses of order $\mathcal{M}_c \sim 5\times 10^{-2} M_\odot$, the range of reach of the search is of order $1$~Mpc, while for chirp masses of order $\mathcal{M}_c \sim 6\times 10^{-4} M_\odot$, the range of reach of the search is of order $10$~kpc.

This can be compared with the sight distance obtained using a fully coherent search, which is given by~\cite{Maggiore_Vol1}
\begin{align}
    d_{L}^\mathrm{coh} & = \frac{c}{n_\sigma \pi^{2/3}} \sqrt{\frac{2}{15}} \left(\frac{G \mathcal{M}_c}{c^3}\right)^{5/6} \sqrt{\int_{f_\mathrm{min}}^{f_\mathrm{max}} \frac{df}{f^{7/3} S_n (f)}}.
    \label{eq:dL_coh}
\end{align}
If we assume that the coherent search is performed between 20Hz and 2048Hz, with a typical LIGO O3 sensitivity, the sight distance of the coherent search is approximately given by
\begin{align}
    d_{L}^\mathrm{coh} \! = \! 0.544\mathrm{Mpc}\!\! \left(\!\frac{\mathcal{M}_c}{10^{-2}M_\odot}\!\right)^{\!\!5/6} \!\!\! \left(\!\frac{n_\sigma}{10}\!\right)^{\!\!-1}\! \!\! \left(\!\frac{\int_{f_\mathrm{min}}^{f_\mathrm{max}} \frac{df}{f^{7/3} S_n (f)}}{3.85 \times 10^{43} \mathrm{Hz}^{-1/3}}\!\right)^{\!\!\! 1/2}.
    \label{eq:dL_coh_numbers}
\end{align}
This is interpreted as an optimal sensitivity, and we see that the Viterbi algorithm can reach the luminosity distance with the same order of magnitude.

\begin{figure}[!t]
\begin{center}
\includegraphics[width=0.5\textwidth]{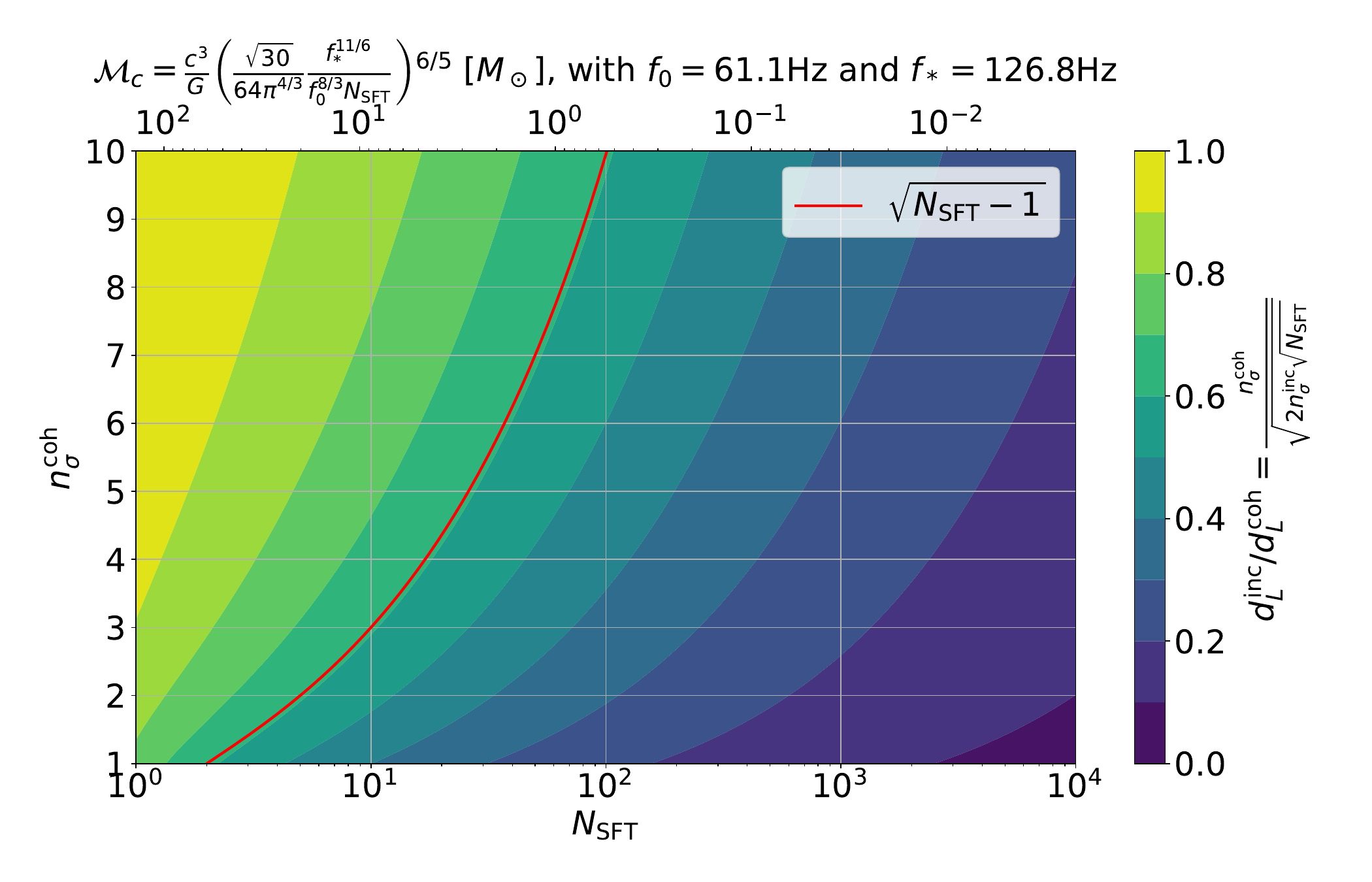}
\end{center} 
\caption{Ratio between the coherent and incoherent sight distances computed, using Eq.~\eqref{eq:dL_inc_coh_ratio} and applying the condition of Eq.~\eqref{eq:equal_FAPs} to impose both to correspond to the same FAP. The contour lines of the ratio are shown as a function of the number of sigmas of the coherent search $n_\sigma^\mathrm{coh}$ and the number of SFTs $N_\mathrm{SFT}$ (or equivalently the chirp mass $\mathcal{M}_c$, which is related via Eq.~\eqref{eq:N_SFT_f0_fs}) and is shown on the upper horizontal axis. The red curve is plotted to show the boundary of $n_\sigma^\mathrm{coh} = \sqrt{N_\mathrm{SFT} - 1}$, below which the distribution of $n_\sigma^\mathrm{inc}$ can be approximated by a Gaussian.
}
\label{fig:dL_inc_coh}
\end{figure}

Comparing the predicted sight distances of the coherent (Eq.~\eqref{eq:dL_coh}) and incoherent (Eq.~\eqref{eq:horizon_distance_TCBC_f0_fs}) cases, it might appear that we could have $d_{L}^\mathrm{coh} < d_{L}^\mathrm{inc}$, contradicting the expectation that a fully-coherent matched-filter search is optimal. Indeed, the ratio between both sight distances is given by
\begin{equation}
    \frac{d_{L}^\mathrm{inc}}{d_{L}^\mathrm{coh}} = \frac{n_\sigma^\mathrm{coh}}{\sqrt{2 n_\sigma^\mathrm{inc}\sqrt{N_\mathrm{SFT}}}} \, ,
    \label{eq:dL_inc_coh_ratio}
\end{equation}
where we have assumed that the coherent and incoherent searches cover the same frequency range, i.e. $f_\mathrm{min} = f_0$ and $f_\mathrm{max} = f_*$ in Eq.~\eqref{eq:dL_coh}, and we have used 
\begin{equation}
    N_\mathrm{SFT} = \frac{T_\mathrm{CBC}}{T_\mathrm{SFT}} = \frac{\sqrt{30}}{64\pi^{4/3}} f_*^{11/6} f_0^{-8/3} \left(\frac{G \mathcal{M}_c }{c^3}\right)^{-5/6} \, .
    \label{eq:N_SFT_f0_fs}
\end{equation}
A priori, one could think that the ratio $d_{L}^\mathrm{inc}/d_{L}^\mathrm{coh}$ of Eq.~\eqref{eq:dL_inc_coh_ratio} can become greater than 1, if $n_\sigma^\mathrm{coh} > \sqrt{2 n_\sigma^\mathrm{inc}\sqrt{N_\mathrm{SFT}}}$. However, we have to take into account that $n_\sigma^\mathrm{coh}$ and $n_\sigma^\mathrm{inc}$ indicate different confidences due to the difference in their distributions, and we should compare them at a fixed confidence. In the case of the coherent search, $n_\sigma^\mathrm{coh}$ follows a normal $\mathcal{N}(0,1)$ distribution, and therefore the false alarm probability (FAP) is
\begin{equation}
    \mathrm{FAP}_\mathrm{coh} = P_{\mathcal{N}(0,1)}(\rho^\mathrm{mf} > n_\sigma^\mathrm{coh}) = \frac{1}{2} \mathrm{erfc} \left( \frac{n_\sigma^\mathrm{coh}}{\sqrt{2}} \right) \, ,
    \label{eq:FAP_coh}
\end{equation}
\noindent where $\mathrm{erfc}(x)$ is the complementary error function. In the incoherent case,$n_\sigma^\mathrm{inc}$ follows a $\chi^2$ distribution with $2 N_\mathrm{SFT}$ degrees of freedom, and therefore:
\begin{align}
     \mathrm{FAP}_\mathrm{inc} & = P_{\chi^2_{2 N_\mathrm{SFT}}}(\rho_\mathrm{tot}^2 > 2N_\mathrm{SFT} + 2 \sqrt{N_\mathrm{SFT}} n_\sigma^\mathrm{inc}) \nonumber \\
     & = Q(N_\mathrm{SFT}, N_\mathrm{SFT} + \sqrt{N_\mathrm{SFT}} n_\sigma^\mathrm{inc}) \, ,
    \label{eq:FAP_coh}
\end{align}
\noindent where $Q(a,x)$ is the regularized upper incomplete gamma function, defined as
\begin{equation}
    Q(a,x)=\frac{1}{\Gamma(a)}\int_{x}^\infty t^{a-1} e^{-t} dt \, .
    \label{eq:gamma_RUI_def}
\end{equation}
When both methods have the same FAP, the values of $n_\sigma$ for the coherent and incoherent searches are related by
\begin{equation}
    \frac{1}{2} \mathrm{erfc} \left( \frac{n_\sigma^\mathrm{coh}}{\sqrt{2}} \right) = Q(N_\mathrm{SFT}, N_\mathrm{SFT} + \sqrt{N_\mathrm{SFT}} n_\sigma^\mathrm{inc}) \, .
    \label{eq:equal_FAPs}
\end{equation}
In Fig.~\ref{fig:dL_inc_coh}, we plot the ratio $d_{L}^\mathrm{inc}/d_{L}^\mathrm{coh}$ by substituting this relation to Eq.~\eqref{eq:dL_inc_coh_ratio}. We observe that $d_{L}^\mathrm{inc}/d_{L}^\mathrm{coh} < 1$ for all plotted values of $N_\mathrm{SFT}$ and $n_\sigma^\mathrm{coh}$, confirming that the coherent search is always more sensitive than the incoherent one. We also plot the line $n_\sigma^\mathrm{coh} = \sqrt{N_\mathrm{SFT} - 1}$, which separates the two different behaviors of the ratio $d_{L}^\mathrm{inc}/d_{L}^\mathrm{coh}$. When $n_\sigma^\mathrm{coh} \ll \sqrt{N_\mathrm{SFT} - 1}$, the $\chi^2_{2 N_\mathrm{SFT}}$ distribution can be well approximated by a Gaussian, and therefore we expect $n_\sigma^\mathrm{inc} \approx n_\sigma^\mathrm{coh}$, and Eq.~\eqref{eq:dL_inc_coh_ratio} reduces to
\begin{equation}
    \frac{d_{L}^\mathrm{inc}}{d_{L}^\mathrm{coh}} \approx \sqrt{\frac{n_\sigma^\mathrm{coh}}{2 \sqrt{N_\mathrm{SFT}}}} .
    \label{eq:dL_inc_coh_ratio_largeN}
\end{equation}
Therefore, in the limit of having a very large number of SFTs, the sight distance of the incoherent search is degraded proportional to $N_\mathrm{SFT}^{-1/4}$ compared to the coherent case. On the opposite limit, when $n_\sigma^\mathrm{coh} \gg \sqrt{N_\mathrm{SFT} - 1}$, we can approximate the ratio $d_{L}^\mathrm{inc}/d_{L}^\mathrm{coh}$ of Eq.~\eqref{eq:dL_inc_coh_ratio} as 
\begin{equation}
    \frac{d_{L}^\mathrm{inc}}{d_{L}^\mathrm{coh}} \approx 1 - \frac{2 N_\mathrm{SFT} - 1}{2(n_\sigma^\mathrm{coh})^2} \log\!\left(\! 1 + \frac{(n_\sigma^\mathrm{coh})^2}{2 N_\mathrm{SFT}} \!\right) \,,
    \label{eq:dL_inc_coh_ratio_smallN}
\end{equation}
\noindent which again confirms that the ratio is always smaller than 1.
In Eq.~\eqref{eq:dL_inc_coh_ratio_smallN} we also observe that $d_{L}^\mathrm{inc}/d_{L}^\mathrm{coh} \approx 1$ when $N_\mathrm{SFT} \ll (n_\sigma^\mathrm{coh})^2$, indicating that in this regime the sensitivity lost by using an incoherent search is small.

\subsection{Constraint on PBH abundance}
The primary motivation for exploring black holes in this mass range is to investigate the possible existence of PBHs. One of our key interests lies in constraining the PBH abundance, often expressed as $f_{\rm PBH}$, the fraction of PBHs relative to the dark matter abundance. Although the constraints heavily depend on the PBH mass function, here, for simplicity, let us consider a monochromatic mass function. Assuming the early binary formation channel~\cite{Sasaki:2016jop}, which is considered to be dominant compared to the late binary formation~\cite{Bird:2016dcv,Clesse:2016vqa,Raidal:2017mfl}, the comoving merger rate density for equal-mass PBHs is given by~\cite{Raidal:2018bbj,Hutsi:2020sol}
\begin{align}
    R=1.6 \times 10^6{\rm Gpc}^{-3}{\rm yr}^{-1} f_{\rm sup} \left(\frac{m_{\rm PBH}}{M_\odot}\right)^{-32/37} f_{\rm PBH}^{53/37} \,,
    \label{eq:rate_theory}
\end{align}
where $m_{\rm PBH}$ is the PBH mass, $f_{\rm sup}$ is the suppression factor that effectively takes into account PBH binary disruptions by early forming clusters due to Poisson fluctuations in the initial PBH separation, by matter inhomogeneities, and by nearby PBHs~\cite{Raidal:2018bbj}. 
The value of the suppression factor is still not clearly understood, but a plausible range is $0.001<f_{\rm sup}<1$ for relatively large PBH fraction $f_{\rm PBH}>0.1$~\cite{Raidal:2018bbj}.

The non-detection of merger events provides an upper limit on the rate density, given by $R_{\rm upper} \sim \langle VT_{\rm obs} \rangle^{-1}$, where $V$ represents the comoving volume to which observations are sensitive. In the range of our interest, specifically $< {\mathcal O}(100)$~kpc, we can safely approximate that the luminosity distance is equivalent to the comoving distance. Assuming the total observation time of $T_{\rm obs}=1$~yr and substituting Eq.~\eqref{eq:dL_numbers} to $V=4\pi d_L^3/3$, we obtain
\begin{align}
    R_{\rm upper}= 53.7 {\rm Mpc}^{-3}{\rm yr}^{-1}  \left(\frac{m_{\rm PBH}}{10^{-2} M_\odot} \right)^{-25/8}\,.
    \label{eq:rete_obs}
\end{align}
By comparing Eqs.~\eqref{eq:rate_theory} and \eqref{eq:rete_obs}, we obtain
\begin{align}
    f_{\rm PBH} =  935 f_{\rm sup}^{-37/53}  \left(\frac{m_{\rm PBH}}{10^{-2} M_\odot} \right)^{-669/424}\,.
\end{align}
While this surpasses the interesting limit, it is important to note that within our galaxy at distances of approximately $50$~kpc, which corresponds to the horizon distances for black hole masses smaller than $\sim 3 \times 10^{-3} M_\odot$, we can anticipate an enhancement in the event rate proportional to the over-density of dark matter. Subsequently, the merger rate density is amplified by a factor of $\sim 3.3 \times 10^5$~\cite{Miller:2020kmv}, resulting in an interesting upper bound of 
\begin{align}
    f_{\rm PBH} =  0.11 f_{\rm sup}^{-37/53}  \left(\frac{m_{\rm PBH}}{10^{-3} M_\odot} \right)^{-669/424}\,.
\end{align}
Here, we have considered only the rate density enhancement within our galactic halo. However, one can also anticipate additional directional enhancements at the galactic center or within the solar system vicinity, which could further strengthen the constraints, as discussed in~\cite{Miller:2020kmv}. 
Note also that this represents the sensitivity curve of O3, and the constraints are expected to improve, yielding an enhanced upper bound with future upgraded detectors.

\subsection{Distribution of the Viterbi detection statistic}
\label{sec:Viterbi_x_dist}
In this subsection, we discuss the distribution of the total SNR of the track recovered by the Viterbi algorithm in detail. The distribution of the total SNR squared $\rho_\mathrm{tot}^2$, defined in Eq.\eqref{eq:tot_SNR}, is described by Eq.\eqref{eq:x_dist} if the track is arbitrarily chosen. However, to be precise, this is not the case because the Viterbi algorithm selects the track with the maximum value of $\rho_\mathrm{tot}^2$ out of all possible tracks allowed by a given transition matrix. Consequently, $\max_{\Vec{\nu}} \rho^2_{\mathrm{tot},\Vec{\nu}} \equiv \rho^2_{\mathrm{tot},\mathrm{max}}$ does not follow the $\chi^2$ distribution of Eq.~\eqref{eq:x_dist}.

Given a flat transition matrix that enables to jump to $n_J$ frequency segments away per time step (where $n_J=1$ is the setting in our case), the probability to calculate Eq.~\eqref{eq:pnu} is given by

\begin{equation}
    p(\nu_i|\nu_{i-1})= 
    \begin{cases}
       \frac{1}{2 n_J + 1} &\; \frac{1}{\Delta f} |\nu_i - \nu_{i-1}| \leq n_J\\
       0 &\; \frac{1}{\Delta f} |\nu_i - \nu_{i-1}| > n_J\\
     \end{cases} \, .
    \label{eq:constant_transition_matrix}
\end{equation}
\noindent where $\Delta f = 1/T_\mathrm{SFT}$ is the frequency resolution of the SFT. There are $2 n_J + 1$ possible ways at each time step, and this is raised to the power of $N_{\rm SFT}$ when considering combinations across all time steps. Then, as the algorithm equally explores all initial frequencies, the total number of tracks that the Viterbi algorithm maximizes over is obtained by multiplying the number of frequency segments $n_f = T_\mathrm{SFT} (f_\mathrm{max} - f_\mathrm{min})$, 
\begin{equation}
    N_t = n_f (2 n_J + 1)^{N_\mathrm{SFT}} \, .
    \label{eq:N_tracks}
\end{equation}
For typical values of $n_J = 1$ and $N_\mathrm{SFT} \sim 1000$, the number of possible tracks becomes huge, of order $\sim 10^{500}$.

If we assume the tracks to be independent, computing the distribution of the maximum total SNR squared among these tracks is simple (see Appendix~\ref{sec:Viterbi_x_dist_uncorr}). However, the correlations between tracks are crucial. Given that different SFT segments are uncorrelated, we have 
\begin{equation}
    \mathrm{Cov}(|\rho^\mathrm{mf}_{ij}|^2, |\rho^\mathrm{mf}_{kl}|^2) = 4 \delta_{ij, kl} \, ,
    \label{eq:rho_ij_cov}
\end{equation}
where $\rho^\mathrm{mf}_{ij}$ denotes the matched filter SNR in the $j$-th frequency of the $i$-th SFT segment, which are distributed as uncorrelated complex Gaussians. Using Eq.~\eqref{eq:rho_ij_cov}, we find that the correlation between two different tracks is given by
\begin{equation}
    r(\rho^2_{\mathrm{tot},\Vec{\nu}_1}, \rho^2_{\mathrm{tot},\Vec{\nu}_2}) \equiv \frac{\mathrm{Cov}(\rho^2_{\mathrm{tot},\Vec{\nu}_1}, \rho^2_{\mathrm{tot},\Vec{\nu}_2})}{\sqrt{\mathrm{Var}(\rho^2_{\mathrm{tot},\Vec{\nu}_1})\mathrm{Var}(\rho^2_{\mathrm{tot},\Vec{\nu}_2})}} = \frac{n_c}{N_\mathrm{SFT}} \,,
    \label{eq:path_correlation}
\end{equation}
where $n_c$ is the number of segments that tracks have in common. 

\begin{figure}[!t]
\begin{center}
\includegraphics[width=0.5\textwidth]{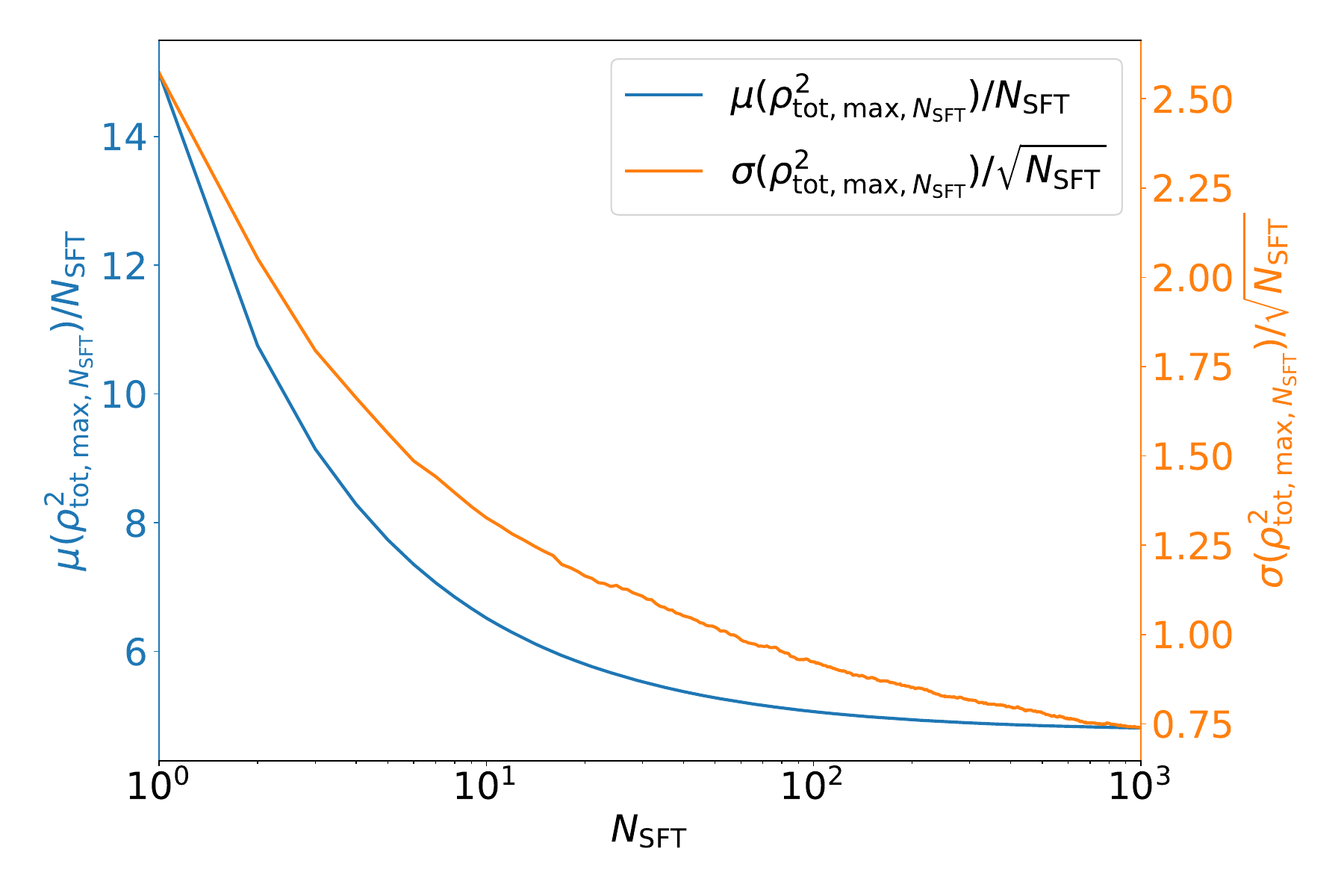}
\end{center} 
\caption{Mean $\mu$ and standard deviation $\sigma$ of the distribution of the maximum over Viterbi tracks of the total SNR $\rho^2_{\mathrm{tot},\mathrm{max}}$ as a function of the number of SFTs, $N_\mathrm{SFT}$. Note that the vertical axis has different scales for the mean (left) and the standard deviation (right). We assume $n_f=1000$ and $n_J=1$, consistent with the values used in our analysis of Sec.~\ref{sec:results}.} 
\label{fig:Viterbi_sigma_mu}
\end{figure}

Given the complexity of the problem, in practice, we investigate the distribution of $\rho^2_{\mathrm{tot},\mathrm{max}}$ using simulated background data assuming Gaussian noise. In Fig.~\ref{fig:Viterbi_sigma_mu}, we plot the mean and standard deviation of this distribution as a function of the number of SFTs, $N_\mathrm{SFT}$ considered in the analysis, while fixing $n_f=1000$ and $n_J = 1$, which are the values consistent with our mock data analysis in the subsequent section. In the simulations, we observe that the distribution depends very weakly on $n_f$, similarly to what happens in the uncorrelated case discussed in Appendix~\ref{sec:Viterbi_x_dist_uncorr}. This can be expected writing Eq.~\eqref{eq:N_tracks} as $N_t =(2 n_J + 1)^{N_\mathrm{SFT} + \log(n_f)}$, and observing that usually $\log(n_f) \ll N_\mathrm{SFT}$.

In Fig.~\ref{fig:Viterbi_sigma_mu}, we find that, similarly to 
Eq.~\eqref{eq:mu_and_sigma_x_dist}, in the absence of signal ($\rho^\mathrm{opt}_\mathrm{tot} = 0$), the mean is proportional to $N_\mathrm{SFT}$ and the standard deviation is proportional to $\sqrt{N_\mathrm{SFT}}$. The proportionality constants are, however, very different, 
and we find
\begin{subequations}
\label{eq:Viterbi_mu_and_sigma_x_dist}
\begin{align}
    \mu(\rho^\mathrm{opt}_\mathrm{tot} = 0, N_\mathrm{SFT}) & = 4.8 N_\mathrm{SFT}  \, , \label{eq:Viterbi_mu_and_sigma_x_dist:mu}\\
    \sigma(\rho^\mathrm{opt}_\mathrm{tot}=0, N_\mathrm{SFT}) & = 0.74 \sqrt{N_\mathrm{SFT}} \, . \; \label{eq:Viterbi_mu_and_sigma_x_dist:sigma}
\end{align}
\label{eq:Viterbi_mu_and_sigma_x_dist:sigma}
\end{subequations}

\begin{figure}[!t]
\begin{center}
\includegraphics[width=0.5\textwidth]{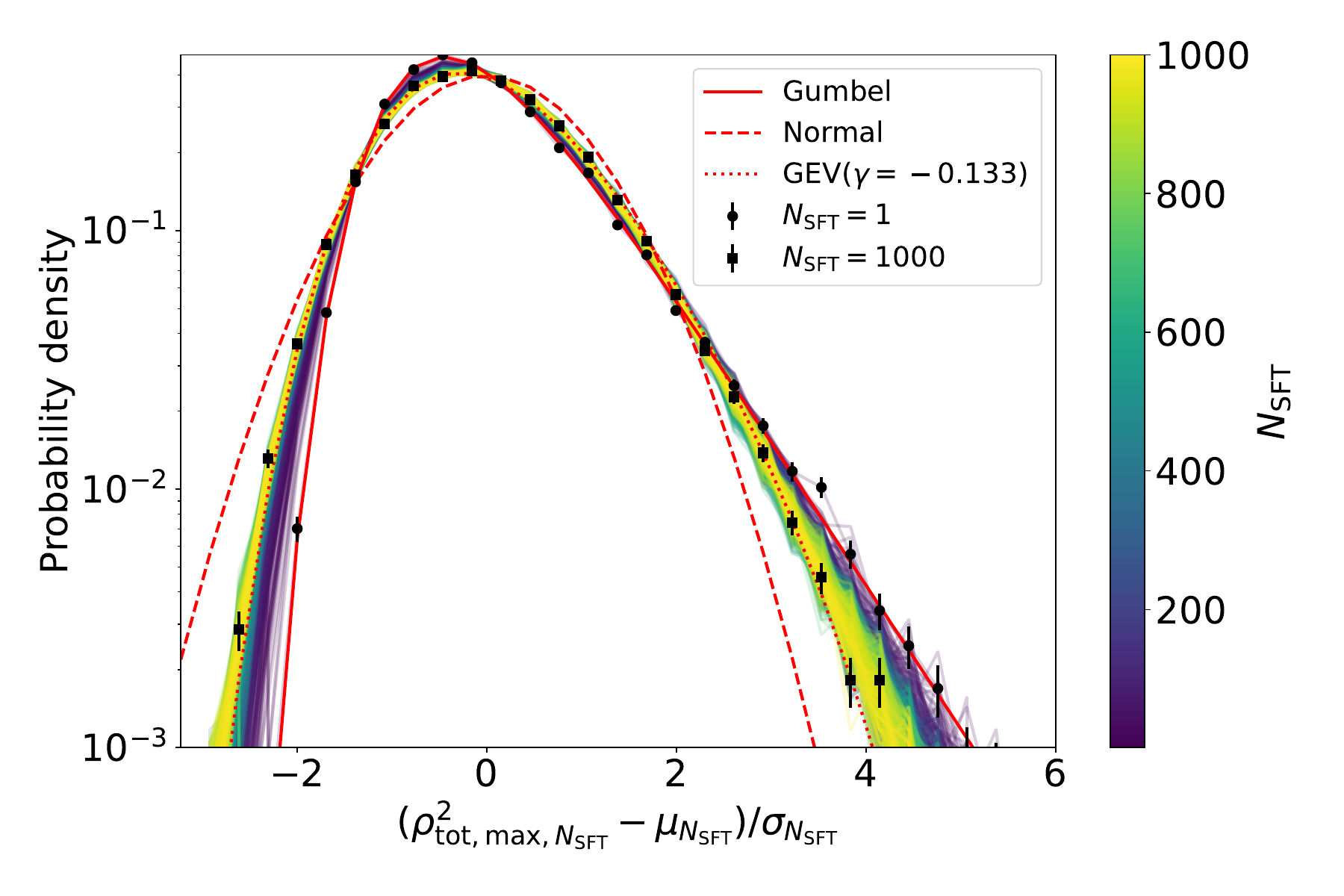}
\end{center} 
\caption{PDF of $\rho^2_{\mathrm{tot},\mathrm{max}}$ scaled and shifted to have zero mean and a standard deviation of 1.} 
\label{fig:Viterbi_nt_distributions}
\end{figure}

In Fig.~\ref{fig:Viterbi_nt_distributions}, we show how the probability distribution function (PDF) of $\rho^2_{\mathrm{tot},\mathrm{max}}$ varies as a function of $N_\mathrm{SFT}$. We can observe that for $N_\mathrm{SFT} = 1$, it resembles a Gumbel distribution, as expected from Appendix~\ref{sec:Viterbi_x_dist_uncorr}, since in this case there is no correlation between paths. However, as $N_\mathrm{SFT}$ increases, the distribution becomes more symmetric, resembling something more similar to a Generalized Extreme Value (GEV) distribution
\begin{equation}
    P_\mathrm{GEV}(x) = \exp\left\{ -(1 + \gamma x)^{-1/\gamma} \right\} \, ,
    \label{eq:GEV_distribution}
\end{equation}
\noindent which approaches the Gumbel distribution in the limit of $\gamma \to 0$. 
Fitting the GEV distribution to the $N_\mathrm{SFT}=1000$ data, we obtain a best fit value of $\gamma = -0.133$, shown also in Fig.~\ref{fig:Viterbi_nt_distributions}.

Similarly to what is done with the critical ratio (CR) of the Hough transform~\cite{Miller:2018rbg}, to preselect events when using the Viterbi algorithm, we can look at the number of standard deviations from the mean the observed total SNR of the reconstructed track is at
\begin{equation}
    n_\sigma = \frac{\rho^2_{\mathrm{tot},\mathrm{max}} - \mu(\rho^\mathrm{opt}_\mathrm{tot} = 0, N_\mathrm{SFT})}{\sigma(\rho^\mathrm{opt}_\mathrm{tot} = 0, N_\mathrm{SFT})} \, .
    \label{eq:Viterbi_n_sigma}
\end{equation}
To evaluate the expectation value of $n_\sigma$, we can use the fact that, for a sufficiently loud signal, the Viterbi track corresponds to the true track, and in this case, $\mu(\rho^\mathrm{opt}_\mathrm{tot}, N_\mathrm{SFT})$ is given by Eq.~\eqref{eq:mu_and_sigma_x_dist:mu}. Therefore, we obtain
\begin{align}
    \langle n_\sigma \rangle & \approx \frac{2 N_\mathrm{SFT} + (\rho^\mathrm{opt}_\mathrm{tot})^2 - \mu(\rho^\mathrm{opt}_\mathrm{tot} = 0, N_\mathrm{SFT})}{\sigma(\rho^\mathrm{opt}_\mathrm{tot} = 0, N_\mathrm{SFT})} \, .
    \label{eq:avg_Viterbi_n_sigma}
\end{align}
Substituting the approximations of $\mu$ and $\sigma$ shown in Eq.~\eqref{eq:Viterbi_mu_and_sigma_x_dist}, for the $n_J = 1$ case, $\langle n_\sigma \rangle$ is approximately given by

\begin{align}
    \langle n_\sigma \rangle & \approx \frac{(\rho^\mathrm{opt}_\mathrm{tot})^2 - 2.88 N_\mathrm{SFT}}{0.74 \sqrt{ N_\mathrm{SFT}}} \, .
    \label{eq:avg_Viterbi_n_sigma_approx}
\end{align}

We expect that, in the absence of signal $(\rho^\mathrm{opt}_\mathrm{tot})^2 \to 0$, we would have $\langle n_\sigma \rangle \to 0$. However this is not the behaviour we observe in Eq.~\eqref{eq:avg_Viterbi_n_sigma} because of the loud signal assumption. We see $\langle n_\sigma \rangle$ vanishes when $(\rho^\mathrm{opt}_\mathrm{tot})^2 = \mu(\rho^\mathrm{opt}_\mathrm{tot}=0,N_\mathrm{SFT}) - 2 N_\mathrm{SFT}$, and we will see in the next section that, at this point, Eq.~\eqref{eq:avg_Viterbi_n_sigma} no longer serves as a good approximation.
The approximation is broken because, as the signal becomes fainter, we do not expect the Viterbi algorithm to recover the true track but a noise-dominated track.

Note that in the derivation of the optimum length of the SFT, done in Sec.~\ref{sec:opt_TSFT_constw}, we did not consider the trials factor of the Viterbi algorithm, and therefore the form of $n_\sigma$ in Eq.\eqref{eq:n_sigmas}, which is used to optimize the SFT length, is much simpler. However, in the large SNR limit in which a signal can be detected and where Eq.~\eqref{eq:avg_Viterbi_n_sigma_approx} applies, we do not expect the trials factor to be so relevant, since we find the same dependence of $n_\sigma \propto (\rho^\mathrm{opt}_\mathrm{tot})^2/\sqrt{N_\mathrm{SFT}}$. Therefore, even if the correction by the trials factor may introduce a factor of difference in the sensitivity estimate, we expect the estimates of Sec.~\ref{sec:opt_TSFT_constw}, in which we have full analytical control, to be broadly applicable.

\begin{figure*}[t!]
\centering
\includegraphics[width=0.495\textwidth]{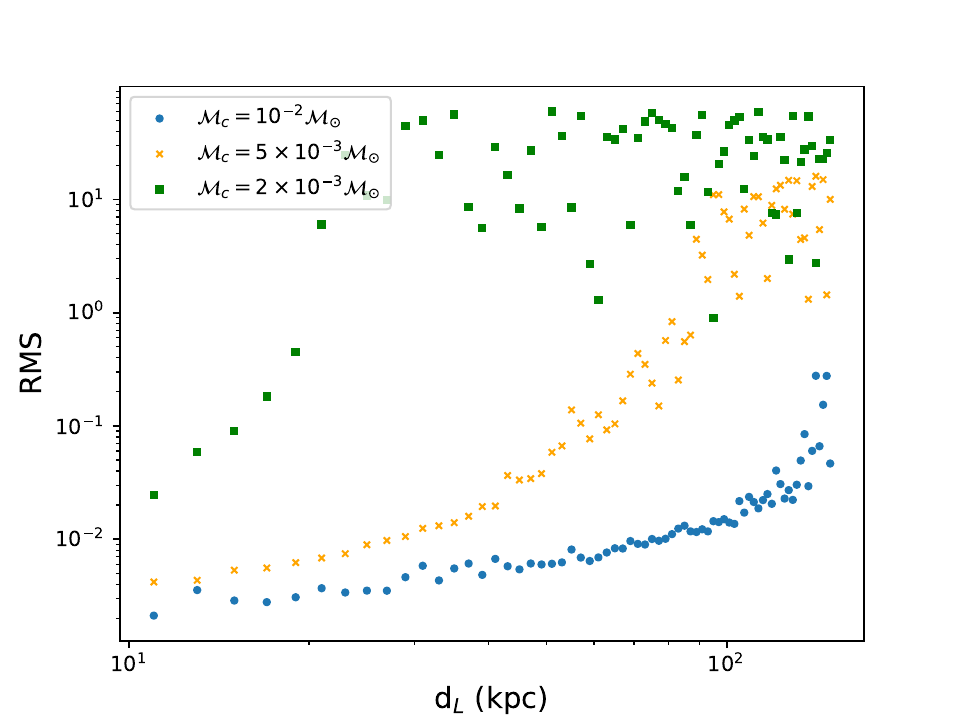}
\includegraphics[width=0.495\textwidth]{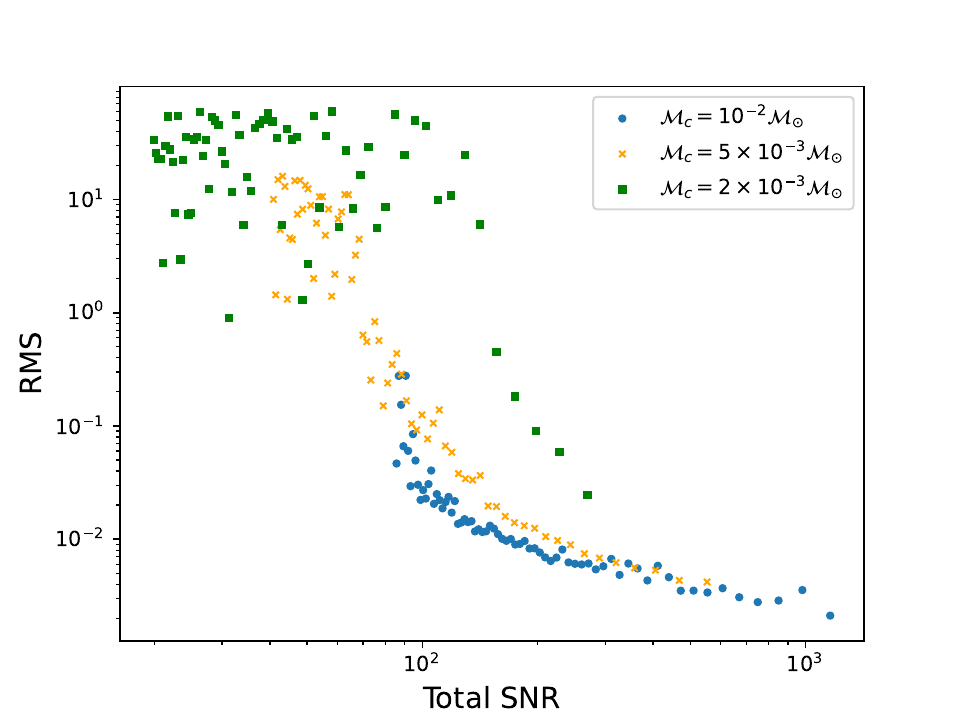}
\caption{\label{fig:rms_dl_SNR}Left panel: The Root Mean Square (RMS) differences between the detected signal by the algorithm and the zero-noise injection as a function of the luminosity distances ($d_L$) of the sources. The RMS is given in units of frequency bins. Right panel: The RMS differences between the detected signal by the algorithm and the zero-noise injection as a function of the \GA{total SNR} for various cases of luminosity distances ($d_L$). }
\end{figure*}

\begin{figure*}[t!]
\centering
\includegraphics[width=0.495\textwidth]{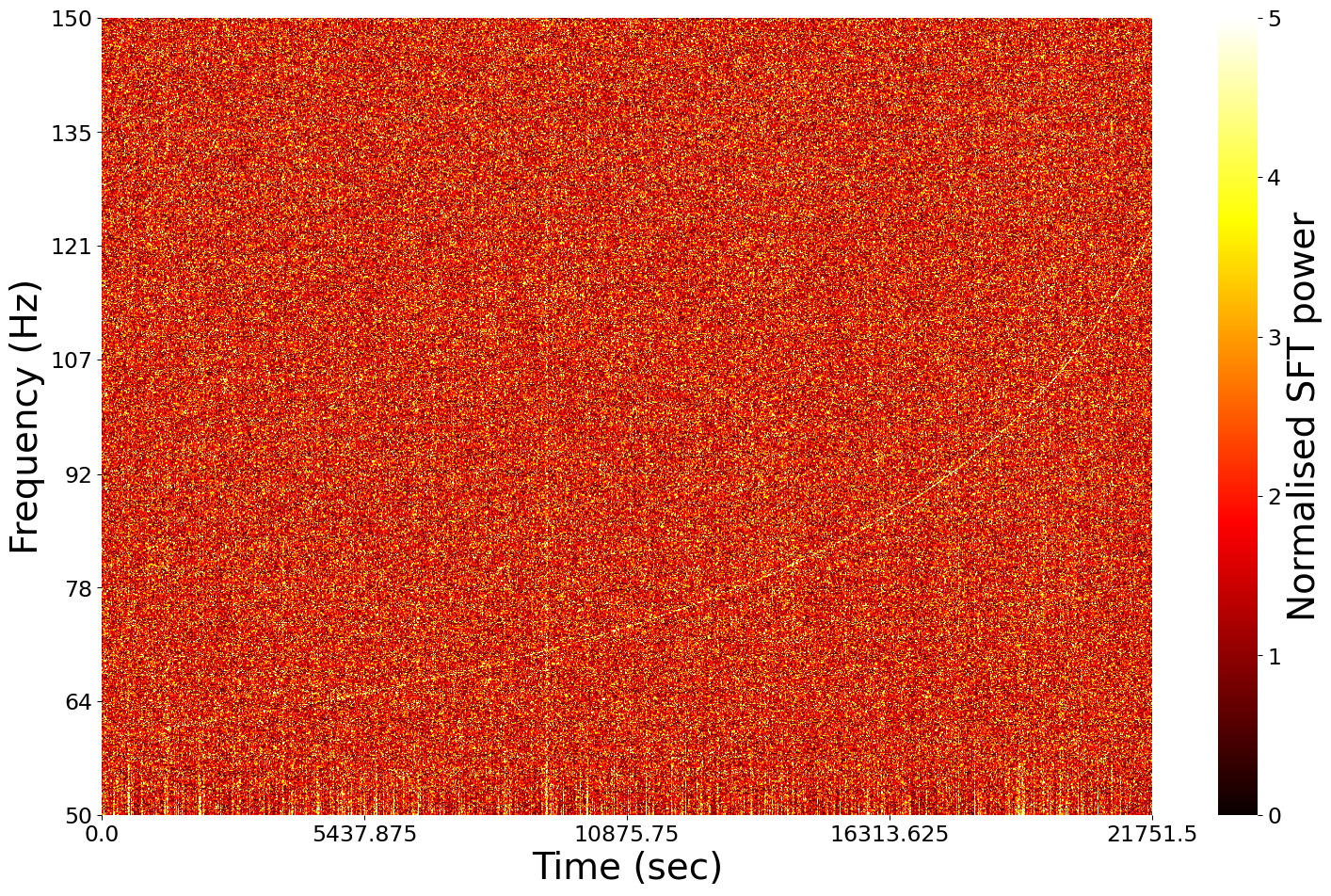}
\includegraphics[width=0.495\textwidth]{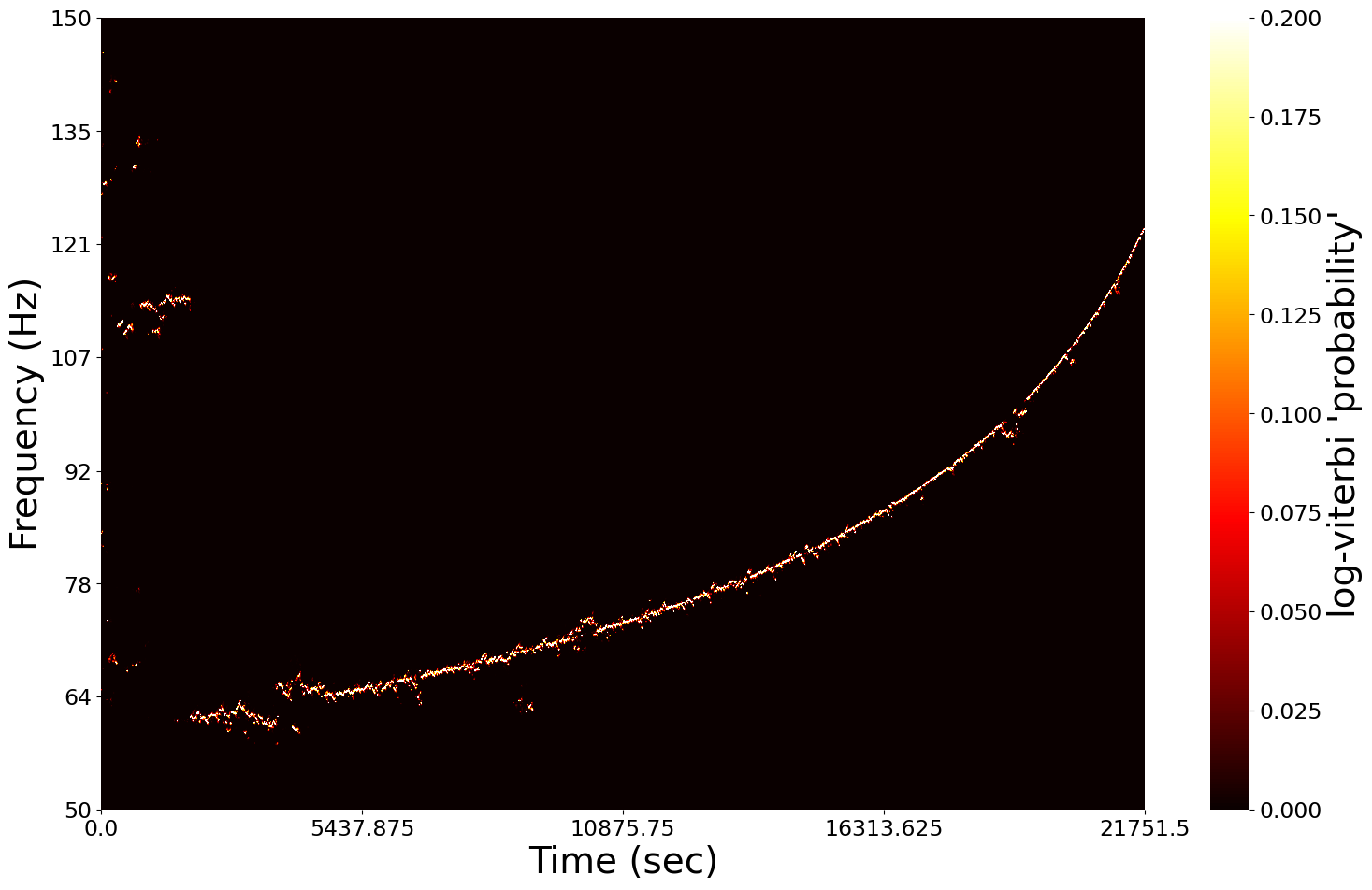}
\caption{\label{fig:signal_spec_det}Signal detection using the Viterbi algorithm in the case of $[\mathcal{M}_c, d_L] = [10^{-2} M_{\odot}, 147 \mathrm{kpc}]$. Left panel: The spectrogram of the injected signal hidden within the Gaussian noise. Right panel: The detected signal by the Viterbi algorithm. 
}
\end{figure*}

\section{Test with mock data}
\label{sec:results}

In this section, we demonstrate a search of PBH compact binary coalescence with the SOAP algorithm. For simplicity, we focus on cases where the two bodies have equal masses and zero spins. We inject signals into Gaussian noise generated with the O3 PSD, considering three different chirp masses, \ie  $[10^{-2}, 5\times10^{-3}, 2\times10^{-3}] M_{\odot}$.
The CBC signals in the SSB are simulated using the 3.5PN \texttt{TaylorT3} approximant~\cite{TaylorApproximants}. This approximant was chosen, because it gives a closed analytical form of the waveform as a function of time, allowing us to easily compute the signal in segments, which is necessary, due to memory limitations, for preparing the very long signals considered in this study. 
We then compute the time-dependent projection of the signal into the Hanford and Livingston detectors using \texttt{lalsuite}~\cite{lalsuite}.

Following the theoretical predictions for the optimum SFT length given in Eq. \eqref{eq:TSFTopt}, we find that the optimum SFT lengths needed for 
chirp masses considered here are $[8.5, 15, 32]$~s, respectively. We use the \texttt{LALPulsar} library~\cite{lalsuite, swiglal}, more specifically the \texttt{lalpulsar\_MakeSFTs} function, to create the SFTs.

Using the python \texttt{soapcw} library~\cite{soapcw}, we have successfully detected the injected signals that correspond to PBH inspiral signatures. In Fig.~\ref{fig:rms_dl_SNR}, we plot the Root Mean Square (RMS), differences between the injected signal and the track detected by the Viterbi algorithm given in units of frequency bins, as functions of the luminosity distance of the source (left panel) and their \GA{total SNR} (right panel). The larger the RMS value, the larger the deviation between the injected signal and the detected one. For a CBC signal, the power deposited in each SFT bin can be very different (the amplitude of the signal drastically increases as the system evolves). Since we expect the Viterbi algorithm to fit better the parts of the track containing more signal, we have defined the RMS differently to Ref.~\cite{Bayley:2019bcb}, weighting by the optimum SNR of each SFT, i.e.

\begin{equation}
    \mathrm{RMS} = \sqrt{\frac{1}{N_\mathrm{SFT}}\sum_{i=1}^{N_\mathrm{SFT}} \frac{(\rho_i^\mathrm{opt})^2}{(\rho_\mathrm{tot}^\mathrm{opt})^2} \left(\frac{\nu_{0, i} - \nu_{V, i}}{\Delta f}\right)^2} \, ,
    \label{eq:RMS_def}
\end{equation}

\noindent where $\nu_{0, i}$ and $\nu_{V, i}$ denote the frequencies the tracks go through in the $i$-th SFT for the injected signal and the Viterbi track respectively, while $(\rho_i^\mathrm{opt})^2$ denotes the squared optimum SNR in the $i$-th SFT such that $(\rho_\mathrm{tot}^\mathrm{opt})^2 = \sum_{i=1}^{N_\mathrm{SFT}} (\rho_i^\mathrm{opt})^2$.

As seen in the right panel, we observe that the efficiency of the algorithm is getting progressively worse as we lower the SNR. In the left panel, we observe a similar but reverse correlation between the RMS and the $d_L$ of the sources. The result is consistent with our sensitivity estimate, Eq.~\eqref{eq:dL_numbers}, which indicates $d_L=[190, 92, 36]$kpc (with $n_\sigma=10$) for $[10^{-2}, 5\times10^{-3}, 2\times10^{-3}] M_{\odot}$.

Fig.~\ref{fig:signal_spec_det} displays the spectrogram, illustrating the track of the injected signal in Gaussian noise (left panel) and the track detected by Viterbi (right panel) for the case of $[\mathcal{M}c, d_L] = [10^{-2} M_{\odot}, 147 \mathrm{kpc}]$. The corresponding matched filter SNR is $86.8$. This represents the marginal luminosity distance where the RMS starts to increase for larger distances. We observe that the algorithm successfully traces the signal, with the exception of a small portion at lower frequencies due to the lower SNR in that range. If the source is at a closer distance, the track is accurately detected even at low frequencies.

\begin{figure}[!t]
\begin{center}
\includegraphics[width=0.5\textwidth]{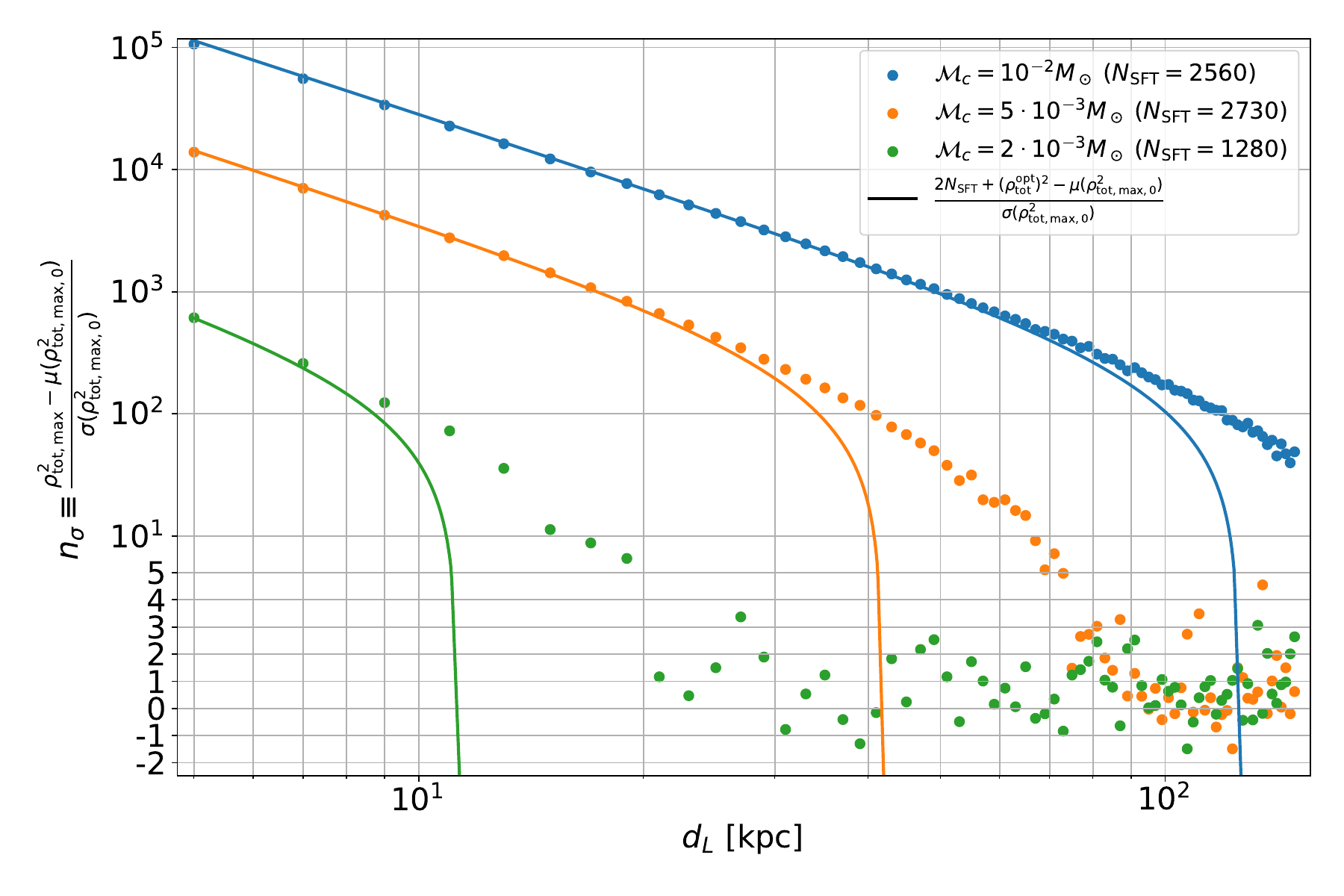}
\end{center} 
\caption{For the different chirp mass cases, we show as a function of luminosity distance $d_L$ the number of standard deviations $n_\sigma$ the SNR is away from the zero signal mean given by Eq.~\eqref{eq:Viterbi_n_sigma} (dots) and approximated expected value of $n_\sigma$ of Eq.~\eqref{eq:avg_Viterbi_n_sigma} (solid lines).} 
\label{fig:Viterbi_n_sigma}
\end{figure}

In Fig.~\ref{fig:Viterbi_n_sigma} we show the number of standard deviations $n_\sigma$, which indicates the extent to which the observed value deviates from the zero signal mean as given in Eq.\eqref{eq:Viterbi_n_sigma}, as a function of the luminosity distance $d_L$ for different chirp masses. The realizations are the same as those in Fig.\ref{fig:rms_dl_SNR}. We observe that the approximation in Eq.\eqref{eq:avg_Viterbi_n_sigma} for the expected value of $n_\sigma$ fits the data well at small values of the luminosity distance, i.e., large SNRs. However, as expected, when $(\rho_\mathrm{tot}^\mathrm{opt})^2 \to \mu(\rho^2_\mathrm{tot, max, 0}) - 2N_\mathrm{SFT}$, the approximation becomes worse and tends to underestimate $n_\sigma$. Comparing with Fig.~\ref{fig:rms_dl_SNR}, we observe that the approximation begins to fail when the RMS becomes large. This implies that the Viterbi algorithm is not following the injected track but rather is tracking a noise-dominated path, and the assumption of following the injected track, used to derive our approximation is no longer valid.
Looking at the distance at which the data points cross $n_\sigma = 10$, we observe that the sight distance derived in Eq.~\eqref{eq:dL_numbers} provides a reliable estimate of the search range. This indicates that, in terms of $n_\sigma$, the Viterbi algorithm does not significantly lose sensitivity due to the trial factor. We have to keep in mind, however, that having the same $n_\sigma$ does not mean having the same statistical significance; this depends on the underlying distribution. The distribution of $\rho^2_\mathrm{tot, max, 0}$, shown in Fig.~\ref{fig:Viterbi_nt_distributions}, has heavier tails than a Gaussian, and therefore the same value of $n_\sigma$ than in Eq.~\eqref{eq:dL_numbers} is less significant.

Lastly, in Tab.~\ref{tab:time_det}, we present the average time required to perform signal detection for each case of $\mathcal{M}_c$ along with the corresponding standard deviation. The calculation was conducted on a standard M1 MacBook Pro, utilizing a sample of detections with varying $\mathcal{M}_c$ values. Note that only the SOAP algorithm part is considered here, and the SFT preparation time is excluded. 

\begin{table}
    \centering
    \begin{tabular}{|c|c|c|c|c|}
    \hline
       $\mathcal{M}_c (M_{\odot})$ & $N_\mathrm{SFT}$ & $n_f$ & $\Delta t (s)$ \\
       \hline
        $10^{-2}$ & 2560 & 552 & 1.34 $\pm$ 0.03 \\
        $5\times 10^{-3}$ & 2730 & 1035 & 1.79 $\pm$ 0.02 \\
        $2\times 10^{-3}$ & 1280 & 2208 & 3.00 $\pm$ 0.01 \\
    \hline
    \end{tabular}
    \caption{The average time ($\Delta t$) needed in order to complete a signal detection using a standard M1 Macbook Pro for each case of $\mathcal{M}_c$ used in our analysis, along with its standard deviation. For reference, we also show the number of SFTs $N_\mathrm{SFT}$ and frequencies $n_f$ for each case. }
    \label{tab:time_det}
\end{table}

The major advantage of the Viterbi algorithm is evident: the search can be efficiently completed in approximately 10 seconds with a laptop computer. 
In fact, the more time-consuming part is the SFT preparation. To prevent signal loss, it is essential to prepare various datasets with differing SFT lengths to search for different PBH masses. Referring to Fig.~\ref{plt:SFT_colormap}, we find that, to maintain a loss of detection efficiency below $10\%$, the SFTs should be prepared with a mass bin size of  $\Delta\log_{10} \mathcal{M}_c \sim 0.3$.

\section{Conclusions}
\label{sec:Conclusions}
In this paper, we have considered the application of Viterbi algorithm to small mass black hole binary search for a range that is not accessible by the current subsolar mass search and CW search. A notable distinction from the CW approach is the necessity to segment the data appropriately, accounting for the gradual evolution of the signal frequency. As a key outcome, we have provided an estimation of the optimal length for the SFT that enhances our search efficiency. Additionally, we have provided sensitivity estimations based on the O3 noise curve. We then demonstrated that it is indeed possible to use the Viterbi algorithm to scan the LVK data and accurately point to the gravitational wave signals. We have tested the method
by injecting signals from sources with 3 different chirp masses, $[10^{-2}, 5\times10^{-3}, 2\times10^{-3}] M_{\odot}$, considering luminosity distances within the range of $d_L = [5, 150]~\mathrm{kpc}$ and successfully captured the signal for close events. 
 
The Viterbi algorithm is highly efficient in terms of computational cost and benefits greatly from the fact that it is agnostic of the characteristic signature of the gravitational waveform of the hidden signal.
On the other hand, it sacrifices in terms of accuracy in its prediction. Therefore, we can use this methodology to provide a quick, first search of possible signals hidden within the data, and then it should be followed by a more thorough, in-depth analysis of the possible targets. 

In the present work, we simulate the data assuming that the noise is stationary and Gaussian. To employ our method in real data analysis, we need to account for the non-stationarity and non-Gaussianity of the detector noise. Particularly, the narrow spectral disturbances, so-called `lines' that can severely degrade the sensitivity~\cite{LSC:2018vzm}. This point remains to be covered in future work. In addition, it would be interesting to cover the cases of asymmetric and extreme mass ratios, as well as to expand the chirp mass range that we consider in our analysis. 
Currently, the primary bottleneck in terms of computation time is the SFT preparation, as we must adjust the FFT length based on the considered chirp mass. We can tackle this challenge by using short FFT
databases (SFDBs)~\cite{Astone:2005fj, Astone:2014esa}, incorporating functions that facilitate easy changes to the FFT length. Moreover, Band Sampled Data (BSD) could potentially offer a solution~\cite{Piccinni:2018akm}. This will be another focus of our future work.

\textit{Numerical Analysis Files:} The codes used by the authors in the analysis of the paper will be made publicly available upon publication \href{https://github.com/GeorgeAlestas/Viterbi_SSM}{here}.

\section*{Acknowledgements}
We would like to thank Andrew Miller for his work reviewing this paper within the LIGO and Virgo Collaborations.
We acknowledge the i-LINK 2021 grant LINKA20416 by the Spanish National Research Council (CSIC), which initiated and facilitated this collaboration. The authors are grateful to J.~Bayley for his useful suggestions regarding the implementation of the SOAP package, and to S.~Clesse for very useful discussions. 
The authors also acknowledge support from the research project PID2021-123012NB-C43 and the Spanish Research Agency (Agencia Estatal de Investigaci\'on) through the Grant IFT Centro de Excelencia Severo Ochoa No CEX2020-001007-S, funded by MCIN/AEI/10.13039/501100011033. GA’s and SK's research is supported by the Spanish Attraccion de Talento contract no. 2019-T1/TIC-13177 granted by the Comunidad de Madrid. 
GM acknowledges support from the Ministerio de Universidades through Grant No. FPU20/02857.
TSY is supported by the JSPS KAKENHI Grant Numbers JP23H04502 and JP23K13099. SK is partly supported by the I+D grant PID2020-118159GA-C42 funded by MCIN/AEI/10.13039/501100011033, the Consolidaci\'on Investigadora 2022 grant CNS2022-135211, and Japan Society for the Promotion of Science (JSPS) KAKENHI Grant no. 20H01899, 20H05853, and 23H00110. 
This material is based upon work supported by NSF's LIGO Laboratory which is a major facility fully funded by the National Science Foundation.

\appendix

\section{Validity of the leading order post Newtonian expansion}
\label{sec:LOPN_val}

One of the key advantages of the Viterbi algorithm is its agnostic nature, enabling us to follow the signal track even when the GW inspiral signal deviates from the power-law spectrum. Here, we evaluate the deviation from the power-law evolution due to PN corrections and show that indeed PN corrections alter the frequency evolution in the cases of our interest.
The next to leading order PN expression is given by~\cite{TaylorApproximants}
\begin{equation}
    f_\mathrm{GW}(t) = \frac{c^3 \theta_c^3}{8 \pi G \mathcal{M}_c} \left( 1 + \left( \frac{743}{2688} + \frac{11}{32} \eta \right) \eta^{-2/5} \theta_c^2 \right) \, ,
    \label{eq:fgw_t_1PN}
\end{equation}
where $\eta = m_1 m_2/(m_1 + m_2)^2$ is the symmetric mass ratio, $\mathcal{M}_c = \eta^{3/5}M$ is the chirp mass and $\theta_c$ is defined as
\begin{equation}
    \theta_c(t) = \left( \frac{c^3(t_c - t)}{5 G\mathcal{M}_c} \right)^{-1/8} \, ,
    \label{eq:theta_c}
\end{equation}
with $t_c$ being the coalescence time. From Eq.~\eqref{eq:fgw_t_1PN}, we can deduce that the correction to the leading order PN approximation is given by
\begin{align}
    \delta f & = f_\mathrm{GW}(t) - f_\mathrm{GW}^{LO}(t) = \frac{c^3 \theta_c^3}{8 \pi G \mathcal{M}_c} \left( \frac{743}{2688 \eta^{2/5}} + \frac{11}{32} \eta^{3/5} \right) \theta_c^2 \nonumber \\
    & = f_\mathrm{GW}^{LO}(t)  \left( \frac{743}{2688 \eta^{2/5}} + \frac{11}{32} \eta^{3/5} \right) \left(\frac{8 \pi G \mathcal{M}_c f_\mathrm{GW}^{LO}(t)}{c^3}\right)^{2/3} \, ,
    \label{eq:fgw_1PN_corr}    
\end{align}
where $f_\mathrm{GW}^{LO}(t)$ is the leading-order PN approximation of the frequency evolution that we want to test. 
The condition that has to be satisfied to neglect higher-order terms is that the error in the frequency approximation is smaller than the resolution of the SFT, that is
\begin{equation}
    \delta f < \frac{1}{T_\mathrm{SFT}} \, .
    \label{eq:condtion_df}
\end{equation}
Using Eq.~\eqref{eq:fgw_1PN_corr}, this is equivalent to
\begin{equation}
    f < \left( \frac{743}{2688 \eta^{2/5}} + \frac{11}{32} \eta^{3/5} \right)^{-3/5} T_\mathrm{SFT}^{-3/5} \left(\frac{8 \pi G \mathcal{M}_c }{c^3}\right)^{-2/5} \, , 
    \label{eq:condition_f}
\end{equation}
where we have defined $f_\mathrm{GW}^{LO}(t) \equiv f$ to simplify the notation. We observe that Eq.\eqref{eq:condition_f} establishes the maximum frequency up to which the leading-order PN approximation is valid. In Eq.\eqref{eq:f_star}, we have also determined that, for a given SFT length, there is a maximum frequency $f_*$ beyond which the energy of the SFT will be contained in a single frequency bin. If in Eq.~\eqref{eq:condition_f} we evaluate $f=f_*(T_\mathrm{SFT})$ to check the condition at the point where it is more likely to be violated, we find
\begin{align}
    T_\mathrm{SFT} & < \frac{1944 \pi^6}{3125} \left( \frac{743}{2688 \eta^{2/5}} + \frac{11}{32} \eta^{3/5} \right)^{-11} \frac{G \mathcal{M}_c}{c^3} \nonumber \\
    & = 0.00467 \mathrm{s} \left(\frac{\frac{743}{2688 \eta^{2/5}} + \frac{11}{32} \eta^{3/5}}{0.6309}\right)^{-11} \left( \frac{\mathcal{M}_c}{10^{-2} M_\odot} \right) \, .
    \label{eq:condtion_TSFT}    
\end{align}
where for the evaluation we have assumed the equal mass case ($\eta=1/4$) which gives the loosest constraint. For typical SFT lengths used in GW searches, which are $\sim$O(1s) in the shortest cases, we observe that there should be measurable deviations from the leading order PN evolution.

Finally, this condition becomes more severe for asymmetric mass case (small $\eta$). Therefore, the advantages of the Viterbi algorithm may become more prominent, particularly in scenarios such as mini EMRI searches~\cite{Guo:2022sdd}.

\section{Distribution of Viterbi total SNR for uncorrelated paths}

\begin{figure}[!t]
\begin{center}
\includegraphics[width=0.5\textwidth]{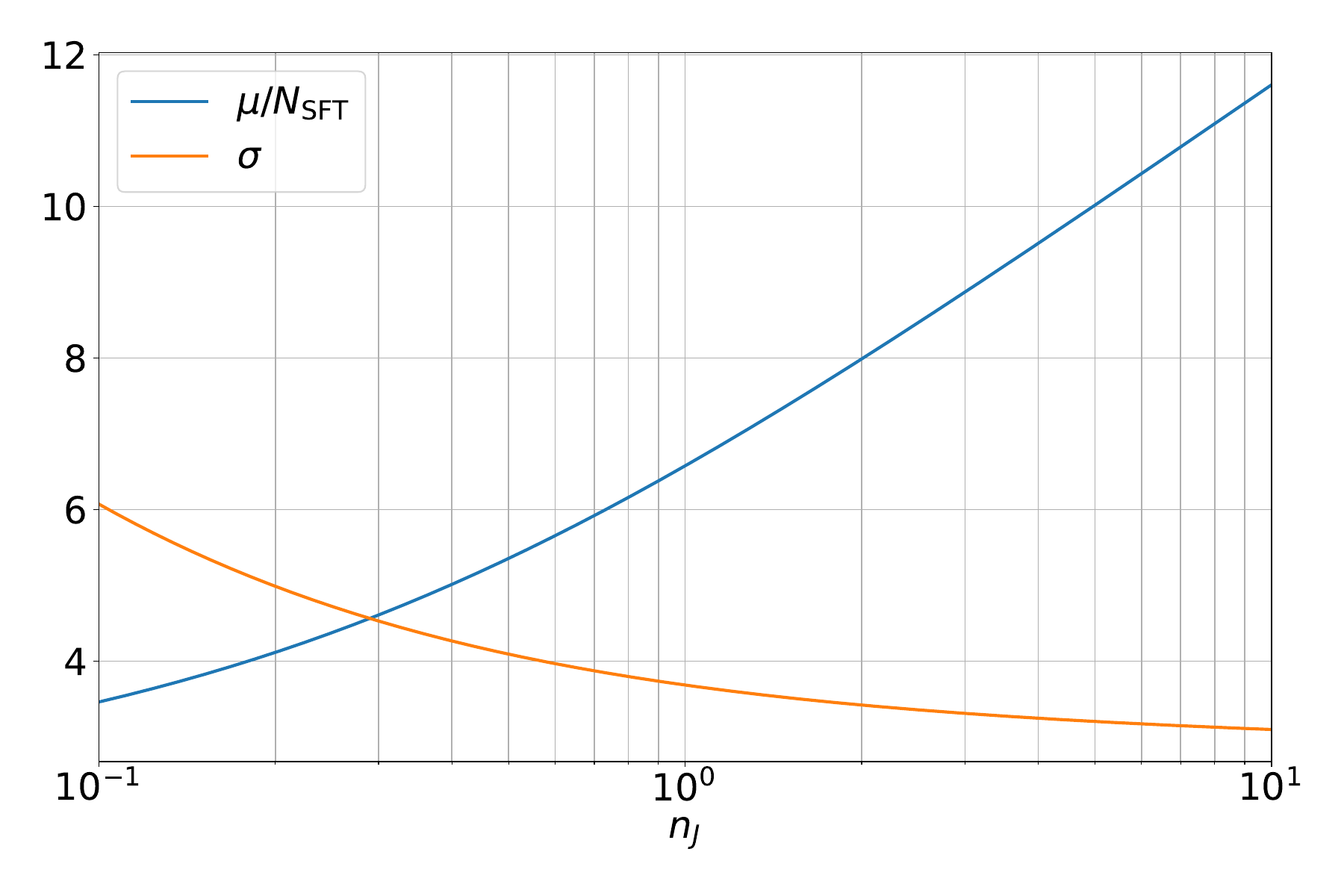}
\end{center} 
\caption{Mean $\mu$ and standard deviation $\sigma$ of the total SNR of the Viterbi track as a function of $n_J$, computed to leading order in $N_\mathrm{SFT}$. We have computed it using Eq.~\eqref{eq:mu_sigma_ViterbiIndep}} 
\label{fig:Viterbi_IndepPaths_musigma}
\end{figure}

Here, we provide an analytical estimation of the distribution of the maximum total SNR squared in the case where tracks are not correlated. The cumulative distribution function (CDF) of the maximum $\rho_{\mathrm{tot},\Vec{\nu}}^2$ of $N_t$ independent paths can be computed as

\label{sec:Viterbi_x_dist_uncorr}
\begin{align}
    P_V(x) & \equiv P\left(\bigcap_{\Vec{\nu}} \rho_{\mathrm{tot},\Vec{\nu}}^2 < x\right) = \prod_{\Vec{\nu}} P\left(\rho_{\mathrm{tot},\Vec{\nu}}^2 < x\right)  \nonumber\\
    & = P\left(\rho_{\mathrm{tot}}^2 < x\right)^{N_t} \, .
    \label{eq:prob_independent}
\end{align}
Limiting ourselves to the case with no signal, then $P\left(\rho_{\mathrm{tot}}^2 < x\right)$ is the CDF of a $\chi^2$ distribution with $2 N_\mathrm{SFT}$ digrees of freedom,
\begin{equation}
    P\left(\rho_{\mathrm{tot}}^2 < x\right) = 1 - \frac{\Gamma(N_\mathrm{SFT}, x/2)}{\Gamma(N_\mathrm{SFT})} \, ,
    \label{eq:CDF_chi2}
\end{equation}
where $\Gamma(N_\mathrm{SFT}, x/2)$ is the upper incomplete gamma function \cite{Abramowitz_and_Stegun}.
Then we obtain
\begin{equation}
    P_V(x) = \exp\left\{ N_t \log\left(1 - \frac{\Gamma(N_\mathrm{SFT}, x/2)}{\Gamma(N_\mathrm{SFT})}\right) \right\} \,.
\end{equation}

Now, our interest lies in the center of the probability distribution rather than its tail. Therefore, when $N_t \gg 1$, the logarithmic component is a negative number which should be very small. In this regime, we can employ the approximation $\log(1-X)\approx -X$ and proceed to obtain

\begin{align}
    P_V(x) 
    & \approx \exp\left\{ - N_t \frac{\Gamma(N_\mathrm{SFT}, x/2)}{\Gamma(N_\mathrm{SFT})} \right\} \nonumber \\
    &= \exp\left\{ -e^{-\alpha(x)} \right\}\, ,
    \label{eq:prob_independent_approx}
\end{align}
\noindent where we have defined
\begin{align}
    \alpha(x) = \log\frac{\Gamma(N_\mathrm{SFT})}{N_t} - \log{\Gamma(N_\mathrm{SFT}, x/2)} \, .
    \label{eq:alpha_gumbel}
\end{align}
In the limit of very large $N_\mathrm{SFT}$, $\alpha(x)$ can be approximated as~\cite{Abramowitz_and_Stegun}
\begin{align}
    \alpha(x) & \approx \log\frac{\Gamma(N_\mathrm{SFT})}{N_t} + \frac{x}{2} - (N_\mathrm{SFT} - 1)\log\frac{x}{2} + O\left(\frac{x}{N_\mathrm{SFT}} \right)\, .
    \label{eq:alpha_gumbel_approx}
\end{align}
This $\alpha(x)$ becomes zero when
\begin{equation}
    x_0 = 2 (N_\mathrm{SFT} - 1) z_0 \, , \; \mathrm{where} \;\; \log(z_0) - z_0 + C = 0 \, ,
    \label{eq:x_0_z_0}
\end{equation}
with $C$ given by
\begin{align}
    C & = \log(N_\mathrm{SFT} - 1) \frac{\log(N_t) - \log(\Gamma(N_\mathrm{SFT}))}{N_\mathrm{SFT} - 1} \nonumber\\
    & \approx 1 + \log(2 n_J + 1) + O\left(\frac{1}{N_\mathrm{SFT}}\right) \, .
    \label{eq:C_z_0}    
\end{align}
Here, we have assumed $N_t = n_f (2n_J + 1)^{N_\mathrm{SFT}}$ with $N_\mathrm{SFT} \gg 1$ and $N_\mathrm{SFT} \gg \ln(n_f)$. Therefore, we have $x_0 \sim O(N_\mathrm{SFT})$, and from Eqs.~(\ref{eq:alpha_gumbel_approx}) and (\ref{eq:x_0_z_0}), we obtain
\begin{align}
    \alpha'(x_0) \approx \frac{1}{2}\left(1  - \frac{1}{z_0}\right) + O\left(\frac{1}{N_\mathrm{SFT}} \right)
    \, ,
    \label{eq:alpha_prime_gumbell_approx}    
\end{align}
and $\alpha''(x_0) \sim O(1/N_\mathrm{SFT}) \to 0$. Therefore, the CDF of Eq.~\eqref{eq:prob_independent_approx} follows the Gumbel distribution
\begin{equation}
    P_V(x) \approx \exp\left\{ -e^{-\alpha'(x_0)(x - x_0)} \right\} \, ,
    \label{eq:prob_independent_gumbel_approx}
\end{equation}
in accordance with the extreme value theorem. This is a well known distribution, having a mean $\mu$ and standard deviation $\sigma$ given by
\begin{equation}
    \mu = x_0 + \frac{\gamma}{\alpha'(x_0)} \quad \mathrm{and,} \quad \sigma = \frac{\pi}{\alpha'(x_0)\sqrt{6}} \, ,
    \label{eq:mu_sigma_ViterbiIndep}
\end{equation}
where $\gamma$ is the Euler-Mascheroni constant. In Fig.~\ref{fig:Viterbi_IndepPaths_musigma}, we show these quantities, to leading order in $N_\mathrm{SFT}$, as a function of $N_\mathrm{SFT}$. For a value of $n_J = 1$, which is the number of jumps we allow for in our Viterbi implementation, from Eq.~\eqref{eq:x_0_z_0}, we have $z_0(n_J=1) = 3.289$, and therefore $\mu_{n_J=1} \sim 6.58 N_\mathrm{SFT}$ and $\sigma_{n_J=1} = 3.69$.

\bibliographystyle{unsrtnat}
\bibliography{Refs}

\end{document}